\documentclass[a4paper,11pt]{article}
\usepackage{graphicx,amssymb,bm,latexsym,color,epsf}
\makeatletter
\@addtoreset{equation}{section}

\makeatother
\newcommand{\bea}{\begin{eqnarray}}
\newcommand{\ena}{\end{eqnarray}}
\newcommand{\vs}[1]{\vspace{#1 mm}}
\newcommand{\hs}[1]{\hspace{#1 mm}}
\renewcommand{\a}{\alpha}
\renewcommand{\b}{\beta}
\renewcommand{\c}{\gamma}
\newcommand{\G}{\Gamma}
\renewcommand{\d}{\delta}

\newcommand{\s}{\sigma}
\renewcommand{\t}{\theta}

\newcommand{\la}{\lambda}
\newcommand{\pa}{\partial}

\newcommand{\nn}{\nonumber\\}
\newcommand{\p}[1]{(\ref{#1})}

\newcommand{\lra}{\leftrightarrow}
\newcommand{\br}{\bar R}
\newcommand{\bg}{\bar g}
\newcommand{\bt}{\bar\tau}

\begin{document}

\begin{titlepage}

\begin{flushright}
KU-TP 060 \\
\end{flushright}

\vs{10}
\begin{center}
{\Large\bf Higher Derivative Gravity and Asymptotic Safety\\
in Diverse Dimensions}
\vs{15}

{\large
Nobuyoshi Ohta\footnote{e-mail address: ohtan@phys.kindai.ac.jp}$^{,a}$
and Roberto Percacci\footnote{e-mail address: percacci@sissa.it}$^{,b,c}$
} \\
\vs{10}
$^a${\em Department of Physics, Kinki University,
Higashi-Osaka, Osaka 577-8502, Japan}

$^b${\em International School for Advanced Studies, via Bonomea 265, 34136 Trieste, Italy}

$^c${\em INFN, Sezione di Trieste, Italy}

\vs{15}
{\bf Abstract}
\end{center}
We derive the one-loop beta functions for a theory of gravity with
generic action containing up to four derivatives.
The calculation is done in arbitrary dimension and on an arbitrary background.
The special cases of three, four, near four, five and six dimensions are discussed in some detail.
In all these dimensions there are nontrivial UV fixed points,
which mean that within the approximations there are asymptotically safe trajectories.
We also find an indication that a Weyl-invariant fixed point exists in four dimensions.
The new massive gravity in three dimensions does not correspond to a fixed point.

\end{titlepage}
\newpage
\setcounter{page}{2}

\section{Introduction}

The empirical success of Einstein's theory of gravity does not imply that
the gravitational action must contain only the Hilbert term,
nor that any other terms must be ``small''.
For example, consider terms quadratic in curvature.
Neglecting indices and setting the cosmological constant to zero,
the Lagrangian will be of the form
\bea
\label{Lsymb}
{\cal L}=m_P^2 R+c R^2\ ,
\ena
where $m_P$ is the Planck mass, ``$R^2$'' stands for generic quadratic combinations
of curvature tensors and $c$ stands for dimensionless couplings.
In the weak field limit we can expand around flat space and write in momentum space
\bea
\label{psymb}
{\cal L}=m_P^2 p^2+c p^4\ .
\ena
The agreement of Einstein's theory with observation puts bounds on the $c$'s: $c\ll (m_P/p)^2$.
The strongest bound comes the highest momenta.
{}From the validity of Newton's law in the millimeter range, one gets the bound $c\ll 10^{62}$.
This ridiculously weak bound is just a consequence of our inability
of testing gravity at high energy, and is another way of seeing how good
the effective field theory of gravity really is \cite{donoghue,burgess}.
If one believes that gravity must be described by some quantum theory at some level,
then the logic of effective field theories leads us to expect that ${\cal L}$
will contain terms quadratic in curvatures, with coefficients $c$ presumably of order unity.
In fact, one expects to find in the action all possible diffeomorphism--invariant terms
constructed with the metric and its derivatives.

In this non-renormalizable effective field theory of gravity, the expansion parameter
is $p/m_P=p\sqrt{G}=\tilde G$, the propagator is given
by the Einstein--Hilbert term and higher derivative terms can be treated as perturbations.
On the other hand, it had been shown long ago that if one uses as propagator
the inverse of (\ref{psymb}), then the theory is perturbatively renormalizable \cite{Stelle1}.
The price one pays is that perturbative unitarity is lost.
More precisely, if the coefficients $c$ are chosen so as to make the theory
renormalizable, there is an unphysical ghost state, and if the coefficients $c$
are such that there is no ghost, the theory is not renormalizable.
Several possible ways out of this dilemma have been proposed, and we will mention some later on.
Here we would like to point out that even when this theory is renormalizable,
it is not UV complete, because Newton's constant is an irrelevant coupling
and $\tilde G$ blows up.
The only way it could be UV complete is if $\tilde G$ tended to a finite limit,
in other words if there was a fixed point (FP).
This is called asymptotic safety~\cite{Weinberg}.

The beta functions for higher derivative gravity in four dimensions have been
calculated at one loop long ago. After some early attempts~\cite{julve,ft1},
the correct beta functions for the dimensionless couplings $c$ have been obtained
by Avramidi and Barvinsky~\cite{Av}.
The calculation has been reviewed and extended to $4-\epsilon$ dimensions
in \cite{BS1,BS2}.
The running of Newton's constant and of the cosmological constant were also derived,
but only taking into account the contribution from the universal
logarithmically divergent terms (or equivalently, in dimensional regularization,
from the simple pole at $D=4$).
The non-universal terms, corresponding to quadratic and quartically divergent terms
in the effective action, were calculated later in \cite{CP,CPR}
and they were shown to lead to a FP for Newton's constant
and for the cosmological constant.
The calculation was repeated using different techniques (and hence with slightly
different results for the non-universal part) in \cite{niedermaier,sgrz}.
Some calculations that go beyond the one-loop approximation have been performed
in \cite{lauscher,bms}, and find also for the $c$ coefficients a finite but nonzero
FP value. The significance of this result is not entirely clear at the moment.

Higher derivative gravity is an interesting subject also in three dimensions,
where Einstein theory is trivial (in the sense that it does not have any propagating
degrees of freedom).
If one adds to the Einstein--Hilbert action a Chern--Simons term,
which contains three derivatives, then it contains a propagating massive graviton
whose helicity depends on the sign of the ``topological mass'' \cite{Top}.
The renormalization group (RG) of this ``topologically massive gravity''
has been studied in \cite{ps}, and its supergravity extension in \cite{ppps}.

These theories, however, violate parity.
More recently it has been shown that the addition of specific four-derivative
terms to three-dimensional gravity (with wrong sign of Einstein--Hilbert term)
makes the theory unitary \cite{BHT1}.
The addition of these terms introduces propagating massive graviton
around flat Minkowski and curved maximally symmetric spacetimes (anti-de Sitter
(AdS) and de Sitter (dS) spacetimes).
In contrast to topologically massive gravity, this so--called ``new massive gravity''
preserves parity. This is very interesting in that we have really dynamical theory of
gravity that is unitary even though higher derivative terms are included.
Various aspects of this theory have been later investigated.
Linearized excitations in the field equations were studied in \cite{liu}.
Unitarity is proven for Minkowski spacetime in \cite{nakasone,deser,gullu},
whereas it is discussed in \cite{BHT2} for maximally symmetric spacetimes.
A complete classification of the unitary theories for the most general action
with arbitrary coefficients of all possible terms is given in \cite{Ohta1}.
The partial result of unitarity condition on the flat Minkowski spacetime
was known for the usual sign of the Einstein theory \cite{nishino}.
Unfortunately, the new massive gravity turned out to
be non-renormalizable though the general theory with arbitrary coefficients for the quadratic
curvature terms is renormalizable \cite{deser,BHT3,Ohta2}.
As in the four--dimensional case, perturbative renormalizability is not
sufficient for UV completeness. For that, a FP is needed.

To establish whether new massive gravity, or some other three--dimensional gravity theory,
has a nontrivial FP, it is necessary to study its beta functions.
It is well known that Einstein gravity, in spite of not having any propagating degrees
of freedom, has such a FP, and it was shown in \cite{ps}
that also topologically massive gravity does.
Much less is known in the case when four--derivative terms are present.
The running of Newton's constant and of the cosmological constant has been
studied in \cite{Ohta3} and it has been shown that it has a FP
with the desired properties, but the running of the four--derivative couplings
has not been studied so far.
One of the main motivations of this paper is to derive the RG flow of
generic four--derivative gravities in three dimensions.
The main questions are whether a FP exists (we will see that it does)
and whether the special ratio of couplings that defines new massive gravity
is stable under RG flow (it is not).

Since the derivation of the beta functions changes little in different dimensions,
we will perform most of the calculations in arbitrary dimension and discuss
in detail only the cases $D=4$ and $D=3$, with additional results for $D=5$ and 6.

This paper is organized as follows. In sect.~2, we summarize the gravity theory we consider
with the Einstein term, cosmological constant and quadratic curvatures,
and give definitions of coupling constants.
In sect.~3, we give quadratic expansion of the action necessary for our calculation
of beta functions.
In sect.~4, we give derivation of the beta functions from functional renormalization group
equation (FRGE) in arbitrary dimensions.
Though we have the results on beta functions for arbitrary dimensions, they are very
complicated and the explicit form is not so illuminating.
So in sects.~5 to 9, we discuss the beta functions and their fixed points explicitly
only for dimensions 4, near 4, 3, 5 and 6, respectively,
and show that the theories have nontrivial UV FPs and are asymptotically safe
in all these dimensions.
By analysing beta functions near four dimensions, we find an indication
that Weyl-invariant FP exists in four dimensions.
We also find that the new massive gravity does not correspond to any FP.
Sect.~10 is devoted to our conclusions.

\section{Four--derivative Gravity}

We will consider actions of the general form
\bea
S=\int d^D x \sqrt{- g} \Big[ \frac{1}{\kappa^2}(\s R - 2 \Lambda) + \a R^2
+\b R_{\mu\nu}^2 + \c R_{\mu\nu\rho\la}^2 \Big] ,
\label{action}
\ena
where $\kappa^2=16\pi G$ is the $D$-dimensional gravitational constant,
$\sigma=\pm1$ is the sign of the Hilbert action, $\Lambda$ is the cosmological constant,
$\alpha$, $\beta$, $\gamma$ are the higher derivative couplings.
It is sometimes more convenient to use a different basis for the higher derivative terms,
namely $R^2$, the square of the Weyl tensor
\bea
C^2 = R_{\mu\nu\a\b}^2-\frac{4}{D-2} R_{\mu\nu}^2+ \frac{2}{(D-1)(D-2)} R^2,
\ena
and the Gauss--Bonnet combination
\bea
E = R_{\mu\nu\a\b}^2- 4 R_{\mu\nu}^2+ R^2,
\ena
which is topological for $D=4$ and vanishes identically for $D=3$.
Then the action has the alternative form
\bea
S = \int d^D x \sqrt{- g} \Big[ \frac{1}{\kappa^2}(\s R - 2 \Lambda)
+ \frac{1}{2\la} C^2 - \frac{1}{\rho} E + \frac{1}{\xi} R^2 +\tau \Box R \Big] ,
\label{action1}
\ena
where
\bea
\lambda = \frac{2(D-3)}{(D-2)(\beta+4\gamma)}, ~~
\rho = \frac{4(D-3)}{(D-2)\beta+4\gamma}, ~~
\xi = -\frac{4(D-1)}{4(D-1)\alpha+D\beta+4\gamma}\ .
\ena
or conversely
\bea
\a = -\frac{1}{\rho}+\frac{1}{\xi} +\frac{1}{(D-1)(D-2)\la}, ~~
\b = \frac{4}{\rho} -\frac{2}{(D-2)\la}, ~~
\c = -\frac{1}{\rho} +\frac{1}{2\la}.
\ena
Note that in $D=3$, $C^2$ and $E$ both vanish identically and the form (\ref{action1})
is not appropriate.
The couplings $\lambda$, $\rho$ and $\xi$ have mass dimension $4-D$.
In dimensions higher than three, it is customary to define the dimensionless combinations
\bea
\omega \equiv -\frac{(D-1)\la}{\xi},~~~
\theta \equiv \frac{\la}{\rho}\ .
\ena
Our conventions are summarized in Appendix A.

The variation of each term gives the field equations:
\bea
\s G_{\mu\nu} +\Lambda g_{\mu\nu} + \kappa^2(\a E_{\mu\nu}^{(1)} + \b E_{\mu\nu}^{(2)}
 + \c E_{\mu\nu}^{(3)})
=0,
\ena
where $G_{\mu\nu}$ is the Einstein tensor and
\bea
E_{\mu\nu}^{(1)} &=& 2RR_{\mu\nu}-2\nabla_\mu \nabla_\nu R
 + g_{\mu\nu}\Big(2\Box R -\frac12 R^2\Big), \nn
E_{\mu\nu}^{(2)} &=& 2R_{\mu\la}R_\nu^\la - 2 \nabla^\la \nabla_{(\mu} R_{\nu)\la}
 + \Box R_{\mu\nu} + \frac12 (\Box R -R_{\la\rho}^2 )g_{\mu\nu}, \nn
E_{\mu\nu}^{(3)} &=& 2R_{\mu\rho\la\s}R_\nu{}^{\rho\la\s}+ 4\nabla_{(\rho}\nabla_{\la)}
R_\mu{}^\rho{}_\nu{}^\la -\frac12 g_{\mu\nu} R_{\rho\la\s\tau}^2.
\ena

\section{Quadratic expansion of the action}

We will apply the standard background field method, expanding the metric as
\bea
g_{\mu\nu}= \bg_{\mu\nu} + h_{\mu\nu}\ .
\label{fluc}
\ena
In order to derive the effective action at the one-loop level,
or to calculate the one-loop beta functions,
we need the expansion of the action to second order in $h_{\mu\nu}$.
This calculation is discussed in detail in Appendix B.
The most complicated part comes from the expansion of the terms quadratic in curvature.
We report here the final form, where we have dropped terms with linear derivatives acting
on the fluctuation and terms with two derivatives acting on a background curvature
(omitting indices, these are of the form $h(\nabla\br)\nabla h$ and $h(\nabla\nabla\br)h$;
such terms do not contribute to the final results, see for example \cite{Av,BS2}).
The terms proportional to $\alpha$ can be written in the form
\bea
&&
\a h^{\mu\nu} \Big[ \nabla_\mu \nabla_\nu \nabla_\a \nabla_\b
- \bg_{\mu\nu} \Box \nabla_\a \nabla_\b - \bg_{\a\b} \nabla_\mu \nabla_\nu \Box
+\bg_{\mu\nu} \bg_{\a\b} \Box^2 -\br \bg_{\nu\b} \nabla_\a \nabla_\mu \nn
&& - (2 \br_{\mu\nu} - \br \bg_{\mu\nu})\nabla_\a \nabla_\b
+ 2 \br_{\mu\nu} \bg_{\a\b}\Box +\frac12 \br (\bg_{\mu\a}\bg_{\nu\b}
- \bg_{\mu\nu}\bg_{\a\b})\Box -\br \br_{\a\b}\bg_{\mu\nu} \nn
&&
+ 2\br \br_{\mu\a} \bg_{\b\nu} + \br_{\mu\nu} \br_{\a\b}
-\frac{1}{4}J_{\mu\nu\alpha\beta}\br^2
\Big] h^{\a\b}\ .
\label{r1}
\ena
Here and in what follows, a bar indicates that the quantity is evaluated on the background;
the indices are raised, lowered and contracted by the background metric $\bg$,
the covariant derivative $\nabla$ is constructed with the background metric.
The tensor $J$ is defined by
\bea
J_{\mu\nu\a\b}=\d_{\mu\nu,\a\b}-\frac12 \bg_{\mu\nu} \bg_{\a\b}
\label{defJ}
\ena
where
\bea
\d_{\mu\nu,\a\b} = \frac12 (\bg_{\mu\a} \bg_{\nu\b} + \bg_{\mu\b} \bg_{\nu\a})
\equiv \hat 1,
\label{id}
\ena
is the identity in the space of symmetric tensors.
We should note that due to the presence of the external factors of $h$,
the expression in the square bracket is automatically symmetrized under the interchanges
$\mu\leftrightarrow\nu$, $\alpha\leftrightarrow\beta$ and
$(\mu,\nu)\leftrightarrow (\alpha,\beta)$.

The $\b$ terms can be written in the form
\bea
\b h^{\mu\nu}\hs{-7}&& \Big[ \frac12 \nabla_\mu \nabla_\nu \nabla_\a \nabla_\b
- \frac12 \bg_{\mu\nu} \nabla_\a \Box \nabla_\b
-\frac12 \bg_{\nu\b} \nabla_\mu \Box \nabla_\a +\frac14( \bg_{\mu\a} \bg_{\nu\b}
+\bg_{\mu\nu} \bg_{\a\b})\Box^2 +\frac12 \br_{\nu\b} \nabla_\a \nabla_\mu \nn
&& -2 \br_\mu^\rho \bg_{\nu\b} \nabla_\rho\nabla_\a
+ \frac32 \bg_{\a\b}\br_{\rho\mu} \nabla^\rho \nabla_\nu
+ \br_{\mu\a\nu\b} \Box +\frac14 (2 \bg_{\mu\a} \bg_{\nu\b}
- \bg_{\mu\nu} \bg_{\a\b}) \br^{\rho\la}\nabla_\rho \nabla_\la
-\frac32 \br^\rho_\b \br_{\rho\mu\nu\a} \nn
&& + \br_{\mu\rho\la\nu} \br_\a{}^{\rho\la}{}_\b
 +\frac12 \bg_{\nu\b} \br_{\mu\rho}\br^{\rho}_\a
- \bg_{\a\b} \br_{\mu\rho} \br^\rho_\nu
-\frac{1}{4}J_{\mu\nu\alpha\beta}\br_{\rho\sigma}^2
\Big] h^{\a\b}\ ,
\label{r2}
\ena
and the terms proportional to $\c$ are
\bea
\c h^{\mu\nu} \hs{-5}&& \hs{-2} \Big[ \nabla_\mu \nabla_\nu \nabla_\a \nabla_\b
+ \bg_{\mu\a} \bg_{\nu\b} \Box^2
-2 \bg_{\nu\b}\nabla_\mu \Box \nabla_\a
- 2 \bg_{\nu\b} \br_{\mu\rho\la\a} \nabla^\rho \nabla^\la
+ 3 \br_{\mu\a\nu\b}\Box \nn
&&
- 4 \br_{\mu \a\rho\nu} \nabla^\rho \nabla_\b
- 4 \bg_{\nu\b} \br_{\rho\mu} \nabla^\rho \nabla_\a
-2\bg_{\nu\b}\br_{\mu\a}\Box
+ \bg_{\mu\a} \bg_{\nu\b} \br_{\rho\la} \nabla^\rho \nabla^\la \nn
&& -2\bg_{\mu\nu} \br_{\rho\a\la\b}\nabla^\la\nabla^\rho
+ 4 \br_{\mu\a} \nabla_\nu \nabla_\b
+ 2 \bg_{\nu\b} \br_{\mu\la\rho\s} \br_\a{}^{\la\rho\s}
-2 \bg_{\nu\b} \br^{\rho\la} \br_{\mu\rho\a\la}
\nn
&&
- \br_{\mu\la\rho\b} \br_\nu{}^{\rho\la}{}_\a
+ 3 \br_{\mu\a}{}^{\rho\la} \br_{\nu\rho\b\la}
-3 \br_{\mu\la\nu\rho} \br_\a{}^{\rho\la}{}_\b
-3 \br^\rho_\a \br_{\mu\b\nu\rho}
\nn
&&
+2 \br_{\mu\a} \br_{\nu\b}
- \bg_{\mu\nu} \br_\a{}^{\rho\la\s} \br_{\b\rho\la\s}
-\frac{1}{4}J_{\mu\nu\alpha\beta}\br_{\rho\sigma\lambda\tau}^2
\Big] h^{\a\b}.
\label{r3}
\ena
It can be checked that when arranged in the Gauss-Bonnet combination
($\gamma=\alpha$, $\beta=-4\alpha$)
and the background metric is maximally symmetric, one obtains a total derivative.
This gives a nontrivial check of the results.

The BRST transformation for the fields is found to be
\bea
\d_B g_{\mu\nu} \hs{-2}&=&\hs{-2} -\d \la [ g_{\rho\nu}\nabla_\mu c^\rho
 + g_{\rho\mu}\nabla_\nu c^\rho
] \equiv -\d\la{\cal D}_{\mu\nu,\rho}c^\rho, \nn
\d_B c^\mu \hs{-2}&=&\hs{-2} -\d\la c^\rho \nabla_\rho c^\mu,~~~
\d_B \bar c_\mu = i \d\la\, B_\mu, ~~~
\d_B B_\mu = 0,
\label{brst}
\ena
which is nilpotent. Here $c^\mu, \bar c_\mu$ and $B_\mu$ are the Faddeev-Popov ghost,
anti-ghost and an auxiliary field, respectively, and $\d\la$ is an anticommuting parameter.
The gauge fixing term and the Faddeev-Popov ghost terms are concisely written as
\bea
{\cal L}_{GF+FP}/\sqrt{-\bg} \hs{-2}&=&\hs{-2} i \d_B [\bar c_\mu Y^{\mu\nu}
(\chi_\nu-\frac{a}{2} B_\nu)]/\d\la \nn
\hs{-2} &=&\hs{-2} - B_\mu Y^{\mu\nu} \chi_\nu
+ i \bar c_\mu Y^{\mu\nu} ( \nabla^\la{\cal D}_{\la\nu,\rho}
+ b \nabla_\nu{\cal D}_{\la,\rho}^\la) c^\rho +\frac{a}{2} B_\mu Y^{\mu\nu} B_\nu \nn
&\simeq& -\frac{1}{2a} \chi_\mu Y^{\mu\nu} \chi_\nu
+ i \bar c_\mu Y^{\mu\nu} [ g_{\nu\rho} \Box +(2b+1)\nabla_\nu \nabla_\rho +\br_{\nu\rho}
] c^\rho,
\label{gfgh}
\ena
where the auxiliary field $B_{\mu}$ is integrated out in the last line.
Here 
\bea
\chi_\mu &\equiv& \nabla^\la h_{\la\mu} + b \nabla_\mu h, \nn
Y_{\mu\nu} &\equiv& \bg_{\mu\nu} \Box+ c \nabla_\mu \nabla_\nu - d \nabla_\nu \nabla_\mu.
\ena
where $a$, $b$, $c$ and $d$ are gauge parameters.
We choose them such that the non-minimal four derivative terms
$\nabla_\mu \nabla_\nu \nabla_\a \nabla_\b$, $\bg_{\mu\nu} \Box\nabla_\a \nabla_\b$
and $\bg_{\nu\b} \nabla_\mu \Box \nabla_\a$ cancel. This leads to the choice~\cite{BS2}
\bea
a= \frac{1}{\b+4\c},~~~
b= \frac{4\a+\b}{4(\c-\a)}, ~~~
c-d = \frac{2(\c-\a)}{\b+4\c}-1.
\ena
In order to simplify the gauge-fixing term, we will further choose $d=1$.
With these choices, the ghost operator is
\bea
\Delta_{gh}=\delta^\mu{}_\nu\Box+(1+2b)\nabla_\mu\nabla^\nu+\br^\mu{}_\nu\ .
\ena
Then, the quadratic terms in the action can be written in the form
$h^{\mu\nu} {\cal K}_{\mu\nu,\a\b} h^{\a\b}$,
where
\bea
{\cal K}=K \Box^2 + D_{\rho\la} \nabla^\rho \nabla^\la + W.
\ena
The explicit forms of the coefficients are
\bea
(K)_{\mu\nu,\a\b} = \frac{\b+4\c}{4} \Big(\bg_{\mu\a}\bg_{\nu\b}
+\frac{4\a+\b}{4(\c-\a)}\bg_{\mu\nu}\bg_{\a\b} \Big),
\ena
\bea
(D_{\rho\la})_{\mu\nu,\a\b} \hs{-2}&=& \hs{-2}
-2\c \bg_{\nu\b}\br_{\a\rho\la\mu}
+4\c \bg_{\rho\nu} \br_{\la\a\mu\b}
+(\b + 3\c) \bg_{\rho\la} \br_{\mu\a\nu\b}
-(2\b+4\c) \bg_{\a\rho}\bg_{\nu\b} \br_{\mu\la} \nn
&& \hs{-2} -2\c \bg_{\nu\b} \br_{\mu\a} \bg_{\rho\la}
+ \b \bg_{\mu\nu}\bg_{\b\rho} \br_{\a\la}
-2\a \bg_{\a\rho} \bg_{\b\la} \br_{\mu\nu}
+2\a \bg_{\mu\nu} \bg_{\rho\la} \br_{\a\b}
+2 \c \bg_{\mu\nu} \br_{\a\rho\la\b} \nn
&& \hs{-2}
+ \Big(\frac{\a}{2}\br +\frac{\s}{4\kappa^2} \Big) (\bg_{\mu\a} \bg_{\nu\b} \bg_{\rho\la}
- \bg_{\mu\nu} \bg_{\a\b} \bg_{\rho\la}
- 2 \bg_{\nu\b} \bg_{\mu\rho} \bg_{\a\la}
+ 2 \bg_{\mu\nu} \bg_{\a\rho} \bg_{\b\la}) \nn
&& \hs{-2}  +2\c \bg_{\nu\rho} \bg_{\b\la} \br_{\mu\a}
+\Big(\frac{\b}{2}+\c \Big) \bg_{\mu\a} \bg_{\nu\b} \br_{\rho\la}
-\frac{\b}{4} \bg_{\mu\nu} \bg_{\a\b} \br_{\rho\la},
\label{der}
\ena
\bea
(W)_{\mu\nu,\a\b} \hs{-2}&=& \hs{-2}
\c \bg_{\nu\b} \br_\mu{}^{\rho\la\s} \br_{\a\rho\la\s}
+ 5\c \br_{\rho\a\mu\la} \br_{\nu\b}{}^{\rho\la}
- \c \br_{\rho\a\mu\la} \br^\rho{}_{\nu\b}{}^\la
+(\b+3\c) \br_{\rho\mu\la\nu} \br^\rho{}_\a{}^\la{}_\b \nn
&& \hs{-2} +(\b+7\c) \br^\rho_\mu \br_{\rho\a\b\nu}
+ \Big(\frac{\b}{2}+2\c \Big) \br_{\mu\a} \br_{\nu\b}
+\Big(\a\br + \frac{\s}{2\kappa^2}\Big)(\br_{\mu\a\nu\b}+
\bg_{\b\nu}\br_{\mu\a} -  \bg_{\mu\nu}\br_{\a\b}) \nn
&& \hs{-2} + \a \br_{\mu\nu} \br_{\a\b}
+ \frac{1}{8} \Big(\a \br^2 +\b \br_{\rho\la}^2 +\c \br_{\rho\la\s\tau}^2
+\frac{1}{\kappa^2}(\s\br-2\Lambda) \Big)
(\bg_{\mu\nu} \bg_{\a\b}-2\bg_{\mu\a} \bg_{\nu\b} ) \nn
&& \hs{-2} +\Big(\frac{5}{2}\b+4\c \Big) \bg_{\nu\b} \br_{\mu\rho} \br^\rho_\a
- \c \bg_{\mu\nu} \br_\a{}^{\rho\la\s} \br_{\b\rho\la\s}
- \b \bg_{\a\b} \br_{\mu\rho} \br^\rho_\nu
- 2\c \bg_{\nu\b} \br^{\rho\la} \br_{\mu\rho\a\la},
\label{cur}
\ena
where we have used the identity~\p{combo1} and dropped terms with two derivatives acting
on a background curvature.

Since the explicit factors of $h$ have been removed, one should now explicitly
perform the symmetrizations $\mu\leftrightarrow\nu$, $\alpha\leftrightarrow\beta$ and
$(\mu,\nu)\leftrightarrow (\alpha,\beta)$.
The latter involves partial integrations,
where we also drop irrelevant terms linear in $\nabla h^{\a\b}$.
After this symmetrization, the first, second and fourth terms as well as the third term
in the third line, and the first in the fourth line in ~\p{der} produce additional terms,
modifying \p{cur} to
\bea
(W)_{\mu\nu,\a\b} \hs{-2}&=& \hs{-2}
\frac32 \c \bg_{\nu\b} \br_\mu{}^{\rho\la\s} \br_{\a\rho\la\s} 
+ 4\c \br_{\rho\a\mu\la} \br_{\nu\b}{}^{\rho\la}
- \c \br_{\rho\a\mu\la} \br^\rho{}_{\nu\b}{}^\la
+(\b+5\c) \br_{\rho\mu\la\nu} \br^\rho{}_\a{}^\la{}_\b \nn
&& \hs{-2} + 6\c \br^\rho_\mu \br_{\rho\a\b\nu}
+ \Big(\frac{\b}{2}+\c \Big) \br_{\mu\a} \br_{\nu\b}
+\Big(\a\br + \frac{\s}{2\kappa^2}\Big)\Big(\frac12\br_{\mu\a\nu\b}+
\frac32\bg_{\b\nu}\br_{\mu\a} -  \bg_{\mu\nu}\br_{\a\b}\Big) \nn
&& \hs{-2} + \a \br_{\mu\nu} \br_{\a\b}
+ \frac{1}{8} \Big(\a \br^2 +\b \br_{\rho\la}^2 +\c \br_{\rho\la\s\tau}^2
+\frac{1}{\kappa^2}(\s\br-2\Lambda) \Big)
(\bg_{\mu\nu} \bg_{\a\b}-2\bg_{\mu\a} \bg_{\nu\b} ) \nn
&& \hs{-2} +\Big(\frac{5}{2}\b+4\c \Big) \bg_{\nu\b} \br_{\mu\rho} \br^\rho_\a
- \c \bg_{\mu\nu} \br_\a{}^{\rho\la\s} \br_{\b\rho\la\s}
- \b \bg_{\a\b} \br_{\mu\rho} \br^\rho_\nu
- (\b+4\c) \bg_{\nu\b} \br^{\rho\la} \br_{\mu\rho\a\la},\nn
\label{curm}
\ena
where the symmetrization $\a\lra\b$ and $\mu\lra\nu$ should still be understood
and we have used the identity~\p{cid}.
As a final step, we factorize the tensor $K$ in the operator ${\cal K}$:
\bea
{\cal K}=K{\cal H}\ ;\qquad
{\cal H}=\Box^2 + V_{\rho\la} \nabla^\rho \nabla^\la +U\ .
\label{hami}
\ena
The form of the coefficients $V_{\rho\lambda}$ and $U$
is complicated and is reported in Appendix~C.
We are now ready to discuss the beta functions in the RG equation.

\section{Derivation of beta functions from functional renormalization group equation}

In the Wilsonian RG, we consider the effective action $\G_k$
describing physical phenomena at momentum scale $k$, which can be regarded as
the lower limit of the functional integration and the infrared cutoff.
The dependence of the effective action on $k$ gives the RG flow,
which can be written as a FRGE~\cite{wetterich} having on the r.h.s.
a trace of functions of the kinetic operators.

Up to this point, we have considered the action in Minkowski space.
In the following derivation of beta functions, we make Euclideanization.
That changes the signs for the Einstein and cosmological terms,
so we should change $\s\to -\s$
and $\Lambda\to -\Lambda$ in our results for the Minkowskian case.
Also in the discussion of fixed points one has to make the couplings
dimensionless by the definitions
\bea
\tilde\alpha=\alpha k^{4-D}\ ;\quad
\tilde\beta=\beta k^{4-D}\ ;\quad
\tilde\gamma=\gamma k^{4-D}\ ;\quad
\tilde\Lambda=\Lambda/k^2\ ;\quad
\tilde G\equiv \kappa^2 k^{D-2}/16\pi,\ .
\ena

In our quadratic action, we have the three operators: ${\cal H}$
acting on the graviton $h_{\mu\nu}$, the ghost operator $\Delta_{gh}$
and the third ghost operator $Y^{\mu\nu}$.
Let us choose cutoffs for the graviton, ghost and third ghost to be functions
of these operators, respectively:
$K R_k({\cal H})$ for the graviton,
$R_k(\Delta_{gh})$ for the ghosts and
$R_k(Y)$ for the third ghost.
The FRGE says that
\bea
\dot\Gamma_k=\frac{1}{2}\mbox{Tr}\frac{\dot R_k({\cal H})}{P_k({\cal H})}
-\mbox{Tr}\frac{\dot R_k(\Delta_{gh})}{P_k(\Delta_{gh})}
-\frac{1}{2}\mbox{Tr}\frac{\dot R_k(Y)}{P_k(Y)},
\ena
where we define $P_k(z)=z+R_k(z)$, and the dot represents the derivative with
respect to $\ln k$.
One can obtain the beta functions of $\Lambda$, $G$, $\alpha$, $\beta$, $\gamma$
by calculating the terms in the r.h.s. proportional to $\int dx\sqrt{g}$,
$\int dx\sqrt{g}\br$, $\int dx\sqrt{g}\br^2$,
$\int dx\sqrt{g}\br_{\mu\nu}\br^{\mu\nu}$.
$\int dx\sqrt{g}\br_{\mu\nu\rho\sigma}\br^{\mu\nu\rho\sigma}$.

In \cite{Ohta3}, the first two terms have been computed in $D=3$
using spectral sums on the sphere.
In order to separate the beta functions of $\alpha$ and $\beta$,
it is not enough to compute the traces on the sphere,
and in more general spaces we do not know the spectrum.
However, we can compute the r.h.s. using the following
general formulas for the trace of a function of an operator.
Calling $\tilde W$ the Laplace transform of $W$, we have for
a differential operator of order $p$ in $D$ dimensions:
\bea
\mathrm{Tr}[W(\Delta)] &=& \sum_n W(\lambda_n)=
\sum_n \int_{0}^{\infty}\!\! ds \, e^{-\lambda_n s} \tilde{W}(s)
=\int_{0}^{\infty}\!\! ds\,\tilde{W}(s)\,\mathrm{Tr}\, e^{-s\Delta}
\nonumber
\\
&=&
\sum_{n=0}^{\infty}B_{2n}(\Delta)
\int_{0}^{\infty}\!\! ds \, \tilde{W}(s) s^{-\frac{D}{p}+\frac{2n}{p}} =
\sum_{n=0}^{\infty} B_{2n}(\Delta)\, Q_{\frac{D-2n}{p}}(W),
\ena
where $B_{2n}$ are the coefficients appearing in the expansion of the heat kernel of the operator
\bea
\mathrm{Tr}\, e^{-s\Delta} =
\sum_{n=0}^{\infty}B_{2n}(\Delta)s^{-\frac{D}{p}+\frac{2n}{p}},
\ena
and the $Q$-functionals are given (for $m>0$) by
\bea
Q_m(W)=\int_{0}^{\infty}\!\! ds \, \tilde{W}(s) s^{-m}
=\frac{1}{\Gamma(m)}\int_{0}^{\infty}\!\! dz z^{m-1} W(z).
\ena
The last form is the more useful one.
(It is obtained more easily going from right to left.
Insert the Laplace expansion of $W$ in the r.h.s.,
exchange the order of the integrations over $s$ and $z$, and then use the integral
representation of the Gamma function.)
For $m=0$, one has $Q_0(W)=W(0)$.

With this formula the FRGE, expanded up to terms quadratic in curvature, is
\bea
\label{master}
\dot\Gamma_k
&=&
\frac{1}{2}B_0({\cal H})Q\left(4,D/4\right)
+\frac{1}{2}B_2({\cal H})Q\left(4,(D-2)/4\right)
+\frac{1}{2}B_4({\cal H})Q\left(4,(D-4)/4\right)
\nonumber
\\
&&-B_0(\Delta_{gh})Q\left(2,D/2\right)
-B_2(\Delta_{gh})Q\left(2,(D-2)/2\right)
-B_4(\Delta_{gh})Q\left(2,(D-4)/2\right)
\nonumber
\\
&&-\frac{1}{2}B_0(Y)Q\left(2,D/2\right)
-\frac{1}{2}B_2(Y)Q\left(2,(D-2)/2\right)
-\frac{1}{2}B_4(Y)Q\left(2,(D-4)/2\right).
\ena
We have to calculate the $Q$-functionals
$Q\left(p,m\right)=Q_m\left(\frac{\dot R_k}{P_k}\right)$, for an operator of order $p$.
For convenience, we choose the cutoff profile \cite{Litim}
$R_k(z)=(k^p-z)\theta(k^p-z)$, where $z$ is a differential operator of order $p$.
Define $z=yk^p$ and then we have
\bea
R_{k}(z)&=&(k^p-z)\theta (k^p-z) = k^p (1-y)\theta(1-y), \\
\dot{R}_{k}(z)&=&pk^p\theta (k^p-z) = pk^p\theta(1-y), \\
P_k (z)&=& z+R_{k}(z) = k^p\quad \mbox{for }\ z<k^p, \\
\frac{\dot R_k}{P_k}&=&p\,\theta(1-y).
\ena
For $m>0$, we find
\bea
Q(p,m) &=&
\frac{1}{\Gamma(m)}\int_{0}^{\infty}\!\! dz z^{m-1} \frac{\dot{R}_{k}(z)}{P_k (z)}
= \frac{k^{mp}}{\Gamma(m)}\int_{0}^{\infty}\!\! dy y^{m-1} p\theta(1-y)
\nonumber \\
&=&  \frac{pk^{mp}}{\Gamma(m)}\int_0^1\! dy y^{m-1}
=  \frac{pk^{mp}}{m\Gamma(m)}.
\label{charlie}
\ena
Furthermore $Q(p,0)=p$.
In $D=3$, we will need also
\bea
Q(p,-1/2)=\frac{p}{\sqrt{\pi}k^{p/2}}\ ;\qquad
Q(p,-1/4)=\frac{p}{\Gamma(3/4)k^{p/4}}\ .
\ena
which derive from \p{charlie} by analytic continuation.

Next we list the necessary heat kernel coefficients.
{}From \cite{Gusynin1990}, we have
\bea
B_4({\cal H})
\hs{-2}&=& \hs{-2}
\frac{1}{(4\pi)^{D/2}}\frac{\Gamma(D/4)}{2\Gamma((D-2)/2)}
\int d^Dx\,\mathrm{tr}\Bigl[
\frac{\hat 1}{90} \br_{\rho\la\s\tau}^2 -\frac{\hat 1}{90} \br_{\rho\la}^2
+\frac{\hat 1}{36}\br^2 +\frac{1}{6} {\cal R}_{\rho\la}{\cal R}^{\rho\la} \nn
&& \hs{-7} - \frac{2}{D-2} U - \frac{1}{6(D-2)} (2 \br_{\rho\la}V^{\rho\la}
- \br V^\rho{}_\rho) + \frac{1}{4(D^2-4)} (V^\rho{}_\rho V^\la{}_\la
+ 2 V_{\rho\la} V^{\rho\la}) \Bigr],~~
\label{grav}
\ena
where $\hat 1$ is the identity defined in \p{id} and
${\cal R}_{\rho\la}$ is the commutator of the covariant derivatives acting
on the tensor $h^{\a\b}$: ${\cal R}_{\rho\la} = [\nabla_\rho, \nabla_\la ]$.
The partial results for the traces in this formula are collected in Appendix~\ref{traces}.
Collecting, we find
\bea
B_4({\cal H})
\hs{-2}&=&\hs{-2}
\frac{1}{(4\pi)^{D/2}}\frac{\Gamma(D/4)}{2\Gamma((D-2)/2)}
\int d^Dx\, \Biggl[
\br_{\mu\nu\rho\la}^2\left(\frac{D(D+1)}{180}-\frac{D+2}{6}
-\frac{2}{D-2}A_1+\frac{12}{D^2-4}D_1\right)
\nn
&&
-\br_{\mu\nu}^2\left(\frac{D(D+1)}{180}+\frac{2}{D-2}A_2
+\frac{1}{3(D-2)}C_1-\frac{12}{D^2-4}D_2\right)
\nn
&&
+\br^2 \left(\frac{D(D+1)}{72}-\frac{2}{D-2}A_3+\frac{1}{6(D-2)}B_1
-\frac{1}{3(D-2)}C_2+\frac{12}{D^2-4}D_3\right)
\nn
&&
+\frac{\s}{\kappa^2}\br\left(-\frac{2}{D-2}A_4+\frac{1}{6(D-2)}B_2
-\frac{1}{3(D-2)}C_3+\frac{12}{D^2-4}D_4\right)
\nn
&&
+\frac{1}{\kappa^2}\left(-\frac{2\Lambda}{D-2}A_5+\frac{1}{\kappa^2}\frac{12}{D^2-4}D_5\s^2\right)
\Biggr],
\label{fre}
\ena
where the constants $A_i, B_i,C_i$ and $D_i$ are defined in Appendix~\ref{traces}.

{}From \cite{GK1999}, we have for the Euclidean operator
\bea
Y_{\mu\nu}=-\bg_{\mu\nu}\Box+\sigma_Y\nabla_\mu\nabla_\nu+\br_{\mu\nu},
\ena
with $\sigma_Y=1-2\frac{\gamma-\alpha}{\beta+4\gamma}$,
\bea
B_4(Y)
\hs{-2}&=&\hs{-2}
\frac{1}{(4\pi)^{D/2}}\int d^D x \sqrt{-\bg}
\Big[
\frac{D-16+(1-\sigma_Y)^{\frac{4-D}{2}}}{180} \br_{\mu\nu\rho\la}^2
\nn
&& \qquad\qquad
-\frac{D-91+(1-\sigma_Y)^\frac{4-D}{2}}{180} \br_{\mu\nu}^2
+\frac{D-13+(1-\sigma_Y)^\frac{4-D}{2}}{72} \br^2
\Big],
\ena
whereas for the Euclidean ghost operator
\bea
\Delta_{gh\mu\nu}=-\bg_{\mu\nu}\Box+\sigma_g\nabla_\mu\nabla_\nu-\br_{\mu\nu},
\ena
with $\sigma_g=-(1+2b)=-\left(1+2\frac{\beta+4\alpha}{4(\gamma-\alpha)}\right)$,
\bea
B_4(\Delta_{gh})
\hs{-2}&=&\hs{-2}
\frac{1}{(4\pi)^{D/2}}\int d^D x \sqrt{-\bg} \Big[
\frac{D-16+(1-\sigma_g)^{\frac{4-D}{2}}}{180} \br_{\mu\nu\rho\la}^2
\nn
&&
-\frac{(1-\sigma_g )^{-D/2}}{180D(D^2-4)\sigma_g}\Big(
D(D^2-4)\sigma_g^3
-2D(D+2)(D+58)\sigma_g^2
\nn
&&
+(D+2)(D^2+118D+720)\s_g
-1440D
\nn
&&
+(1-\sigma_g )^{D/2}
\{ (D^4-91D^3+596D^2-596D-1440 )\s_g+1440D \}
\Big)\br_{\mu\nu}^2
\nn
&&
+\frac{(1-\sigma_g )^{-D/2}}{72D(D^2-4)\sigma_g}\Big(
D(D^2-4)\sigma_g^3
-2D(D+2)(D+10)\sigma_g^2
\nn
&&
+(D+2)(D^2+22D+144) \s_g
-576
\nn
&&
+ (1-\sigma_g )^{D/2} \{(D^4+11 D^3-28D^2+52D-288)\s_g+576 \}
\Big)
\br^2
\Big],
\ena
The $B_4$ agree with the formulas in \cite{BS2} for $D=4$, but the dependence on $D$ is more
complicated than appears there.

We also need the lower heat kernel coefficients:
\bea
B_2({\cal H})
\hs{-2}&=&\hs{-2}
\frac{1}{(4\pi)^{D/2}}\frac{\Gamma((D-2)/4)}{2\Gamma((D-2)/2)}
\int d^Dx\,\sqrt{-\bg}\,\mathrm{tr}\left[\frac{\br}{6}+\frac{1}{2D}V^\mu_\mu\right],
\\
B_2(\Delta_{gh})
\hs{-2}&=&\hs{-2}
\frac{1}{(4\pi)^{D/2}}\frac{D+5+(1-\sigma_g)^{1-D/2}+\frac{12}{D}((1-\sigma_g)^{-D/2}-1)}{6}
\int d^Dx\,\sqrt{-\bg}\,\br ,
\\
B_2(Y)
\hs{-2}&=&\hs{-2}
\frac{1}{(4\pi)^{D/2}}\frac{D-7+(1-\sigma_Y)^{1-D/2}}{6}
\int d^Dx\,\sqrt{-\bg}\,\br ,
\ena

\bea
B_0({\cal H})
\hs{-2}&=&\hs{-2}
\frac{1}{(4\pi)^{D/2}}\frac{\Gamma(D/4)}{2\Gamma(D/2)}\frac{D(D+1)}{2}
\int d^Dx\,\sqrt{-\bg}\;,
\\
B_0(\Delta_{gh})
\hs{-2}&=&\hs{-2}
\frac{1}{(4\pi)^{D/2}}
(1-\sigma_g)^{-D/2}\left(1+(D-1)(1-\sigma_g)^{D/2}\right)\int d^Dx\,\sqrt{-\bg}\;,
\\
B_0(Y)
\hs{-2}&=&\hs{-2}
\frac{1}{(4\pi)^{D/2}}
(1-\sigma_Y)^{-D/2}\left(1+(D-1)(1-\sigma_Y)^{D/2}\right)\int d^Dx\,\sqrt{-\bg}\;.
\ena

Substituting the heat kernel coefficients in equation (\ref{master}),
and extracting the coefficients of $1$, $\br$, $\br^2$, $\br_{\mu\nu}\br^{\mu\nu}$,
and $\br_{\mu\nu\rho\sigma}\br^{\mu\nu\rho\sigma}$,
we obtain the beta functions of $\Lambda$, $G$, $\alpha$, $\beta$, $\gamma$.
It is possible to write the formulas for general $D$ but they are very complicated
and not especially illuminating.
In the following we shall consider the cases $D=3$, $4$, $5$, $6$ as well as
continuous dimensions close to $4$.

\section{Four dimensions}

\begin{figure}[htp]
\centering
\resizebox{0.4\textwidth}{!}{
\includegraphics{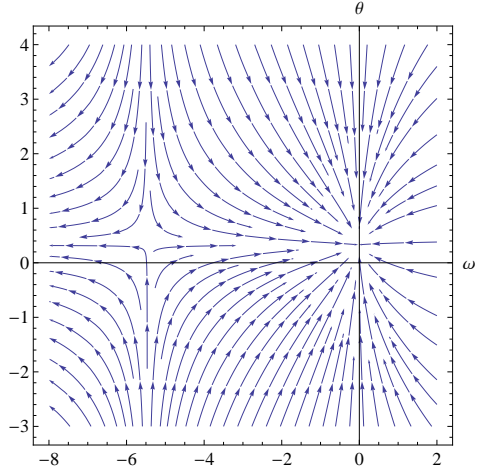}
}
\caption{The flow in the $\omega$-$\theta$ plane in $D=4$
with the fixed points FP$_1$ and FP$_2$.}
\label{f1}
\end{figure}

We begin with the case $D=4$, which is the most studied in the
literature and provides a test of our techniques.
In four dimensions, the couplings $\alpha$, $\beta$ and $\gamma$
(or equivalently $\lambda$, $\xi$ and $\rho$) are all dimensionless.
We use the couplings $\tilde\Lambda=\Lambda/k^2$, $\tilde G=G\,k^2$,
$\lambda$, $\omega$ and $\theta$, as defined in sections 2 and 4.
The beta functions of the three latter couplings form a closed subsystem:
\bea
\beta_{\lambda} & =&  -\frac{1}{(4\pi)^{2}}\frac{133}{10}\lambda^{2},
\\
\beta_{\omega} & =&  -\frac{1}{(4\pi)^{2}}\frac{25 + 1098\,\omega+ 200\,\omega^2}{60}\lambda,
\\
\beta_{\theta} & = & \frac{1}{(4\pi)^{2}}\frac{7(56-171\,\theta)}{90}\lambda\ .
\ena
They agree with the results of \cite{BS2}.
The coupling $\lambda$ has the usual logarithmic approach to asymptotic freedom.
The condition $\lambda=0$ also guarantees the vanishing of the other two beta functions.
In order to find preferred values for $\omega$ and $\theta$, it is customary to define
a rescaled renormalization group time $\bt$ such that $d\bt=\lambda\, dk/k$.
With this variable, one finds
\bea
\frac{d\omega}{d\bt} & =&  -\frac{1}{(4\pi)^{2}}\frac{25 + 1098\,\omega+ 200\,\omega^2}{60}\ ,
\\
\frac{d\theta}{d\bt} & = & \frac{1}{(4\pi)^{2}}\frac{7(56-171\,\theta)}{90}\ .
\ena
which have real FPs
$$
\mbox{FP}_1:(\omega_*,\theta_*)\approx(-5.46714,0.327485)\ ;\qquad
\mbox{FP}_2:(\omega_*,\theta_*)\approx(-0.0228639,0.327485)\ .
$$
This flow is shown in Fig.~\ref{f1}.
Note that $\frac{d\lambda}{d\bt}=-\frac{1}{(4\pi)^{2}}\frac{133}{10}\lambda$,
so also with this variable one sees the FP at $\lambda=0$.

The beta functions for $\tilde G$ and $\tilde\Lambda$ can be written as
\bea
\label{fleq1}
\beta_{\tilde \Lambda} & = &
-2\tilde\Lambda
+p(\lambda,\omega)\tilde\Lambda
-q(\omega)\tilde G \tilde\Lambda
+r(\omega)\tilde G
+s(\lambda,\omega)
+\frac{t(\lambda,\omega)}{\tilde G}\ ,
\\
\beta_{\tilde G} & = & 2 \tilde G-u(\lambda,\omega)\tilde G
-q(\omega)\tilde G^2
\ ,
\label{fleq2}
\ena
where
\bea
p(\lambda,\omega)&=&\frac{1}{(4\pi)^{2}}\frac{1+86\omega+40\omega^2}{12\omega}\lambda,\\
q(\omega)&=&\frac{171+298\omega+152\omega^2+16\omega^3}{36\pi\s (1+\omega)},\\
r(\omega)&=&\frac{283+664\omega+204\omega^2-128\omega^3-32\omega^4}{144\pi(1+\omega)^2},\\
s(\lambda,\omega)&=&-\frac{\s}{(4\pi)^{2}}\frac{1+10\omega}{4\omega}\lambda,\\
t(\lambda,\omega)&=&\frac{\s^2}{(4\pi)^{2}}\frac{1+20\omega^2}{256\pi\omega^2}\lambda^2,\\
u(\lambda,\omega)&=&\frac{1}{(4\pi)^{2}}\frac{3+26\omega-40\omega^2}{12\omega}\lambda.
\ena

To picture the flow of $\tilde\Lambda$ and $\tilde G$,
we set the remaining variables to their FP values $\omega=\omega_*$, $\theta=\theta_*$,
and $\lambda=\lambda_*=0$. Then, defining
$r_*=r(\omega_*)$,
$q_*=q(\omega_*)$,
the flow equations~\p{fleq1} and \p{fleq2} become very simple:
\bea
\beta_{\tilde \Lambda} & = &
-2\tilde\Lambda
+r_*\tilde G
-q_*\tilde G \tilde\Lambda,
\\
\beta_{\tilde G} & = & 2\tilde G-q_*\tilde G^2.
\ena

\begin{figure}[t]
\begin{center}
\begin{minipage}{60mm}
\includegraphics[width=6cm]{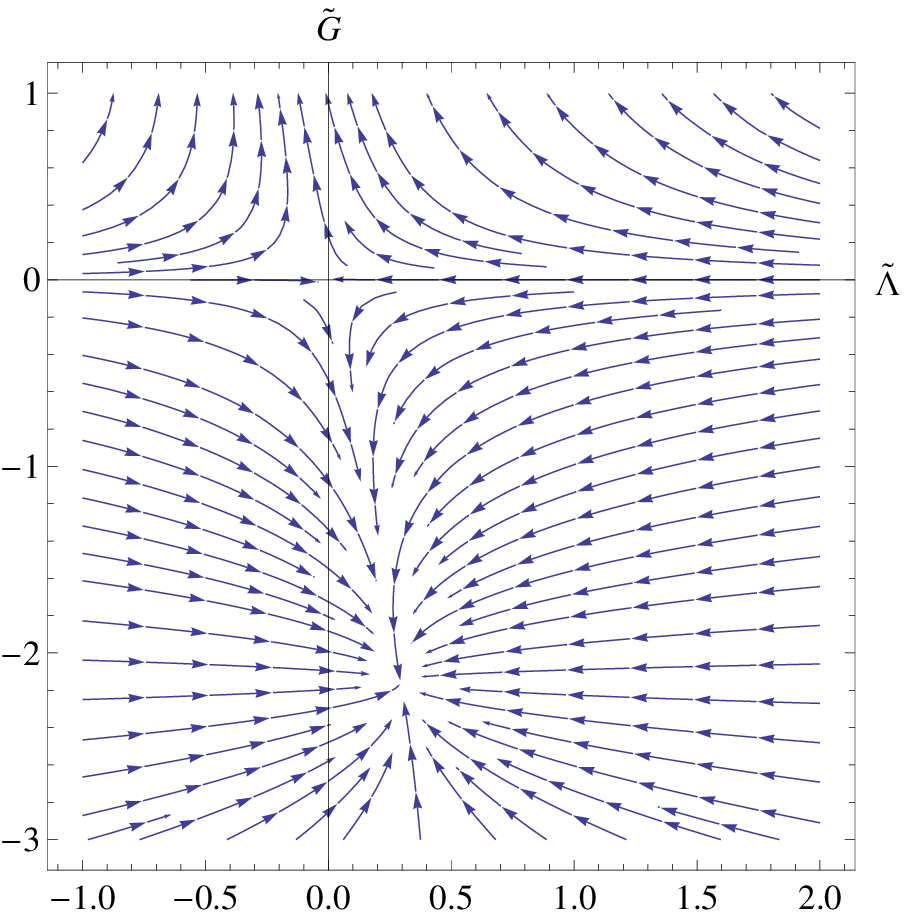}
\put(-90,-15){(a)}
\end{minipage}
\hs{20}
\begin{minipage}{60mm}
\includegraphics[width=6cm]{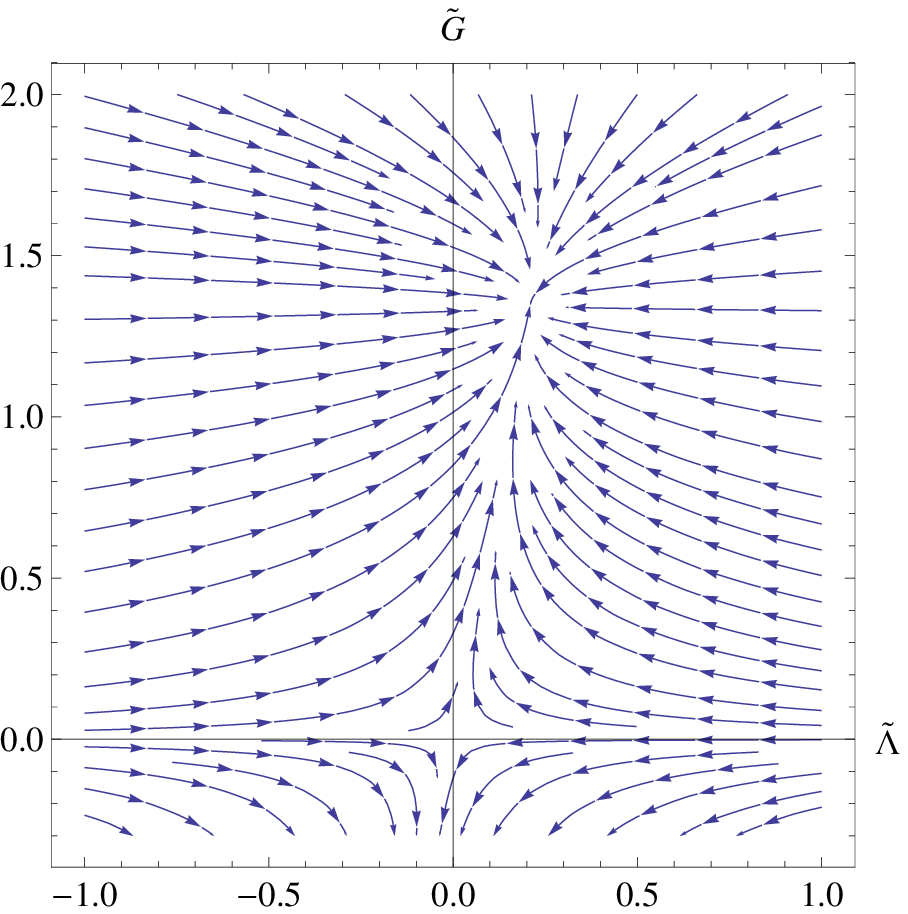}
\put(-90,-15){(b)}
\end{minipage}
\caption{The flow in the $(\tilde\Lambda,\tilde G)$--plane
(a) for the FP $(\tilde{\Lambda}_{*},\tilde{G}_{*})
\approx (0.293, -2.148)$ and
(b) for $(\tilde{\Lambda}_{*},\tilde{G}_{*})
\approx (0.209, 1.346)$.}
\label{fig1}
\end{center}
\end{figure}

The resulting flow in the $(\tilde\Lambda,\tilde G)$--plane is shown in Fig.~\ref{fig1}
for $\s=1$.
It has two FPs for each FP of $\omega$ and $\theta$.
At FP$_1$, we have $r_*\approx-0.545$, $q_*\approx -0.931/\s$.
Then there is the Gaussian FP at $\tilde{\Lambda}=\tilde{G}=0$
and another nontrivial FP at
\bea
\tilde{\Lambda}_{*}=\frac{r_*}{2 q_*}\approx 0.293 \s \ ,\ \ \ \ \
\tilde{G}_{*}=\frac{2}{q_*}\approx -2.148 \s\ .
\ena
At FP$_2$, one has
$r_*\approx 0.620$,
$q_*\approx 1.486/\s$ and again there is a Gaussian FP
and a non-Gaussian one at
\bea
\tilde{\Lambda}_{*}=\frac{r_*}{2 q_*}\approx 0.209 \s \ ,\ \ \ \ \
\tilde{G}_{*}=\frac{2}{q_*}\approx 1.346 \s\ .
\ena
We note that with the normal choice $\sigma=1$, the FP occurs
for positive $G$ in the case of FP$_2$ and for negative $G$ in the case of FP$_1$.

The attractivity properties of these FPs are determined by the stability matrix
$$
M_{ij}=\frac{\partial\tilde{\beta}_{i}}{\partial\tilde{g}_{j}}=
\left( \begin{array}{cc}
-2-q_* \tilde G_* & r_*-q_*\tilde\Lambda \\
0 & 2-2q_* \tilde G_*  \end{array} \right).
$$
At the Gaussian FP the eigenvalues of $M$ are $(-2,2)$;
the attractive eigenvector points along the $\tilde\Lambda$ axis and
the repulsive eigenvector has components $(r_*/4,1)$.
At the non--Gaussian FP the eigenvalues of $M$ are $(-4,-2)$
with the same eigenvectors as before.

Finally we mention that the beta functions of $\lambda$, $\omega$ and $\theta$,
are universal, in the sense that they do not depend on the choice of the cutoff function $R_k$.
This follows from the fact that they are proportional to the universal coefficients $Q(p,0)$.
The beta functions of $\tilde\Lambda$ and $\tilde G$ are not universal.
There are however some contributions that are proportional to $Q(p,0)$
times $B_4$ coefficients, which are universal.
These are the functions $p$, $s$, $t$, $u$, which can be written in terms
of the coefficients given in Appendix~D:
\bea
p&=&\frac{1}{192\pi^2}\left(-12A_4-6A_5+B_2-2C_3+12D_4\right)\ ;
\nonumber\\
s&=&\frac{1}{128\pi^2} B_2\ ;\qquad
t=\frac{1}{512\pi^2}D_5\ ;\nonumber\\
u&=&\frac{1}{192\pi^2}\left(12A_4-B_2+2C_3-12D_4\right)\ .
\ena
On the other hand, the functions $q$ and $r$
are proportional to the scheme--dependent coefficients,
$Q(p,1)$ and $Q(p,2)$, multiplying heat kernel coefficients $B_2$ and $B_0$.
This can be confirmed also by looking at equations (3.2) in \cite{sgrz}.\footnote{
A factor $1+10\omega^2$ in equation (8a) in \cite{CP}
contains a misprint and should read $1+10\omega$.}
It is important to stress that although the value of $Q(p,1)$ and $Q(p,2)$ can be
changed by changing the function $R_k$, they always remain positive,
so that the existence of a nontrivial FP for $\tilde G$ is universal.

\section{Near four dimensions}

Before considering the more interesting cases of three, five and six dimensions,
let us discuss some properties of the theory in $D=4+\epsilon$ dimensions.
This case had been considered previously in \cite{BS2},
but the $D$--dependence of the heat kernel coefficients which we took from
\cite{Gusynin1990,GK1999}
is more complicated than theirs, and so are the beta functions.
Also note that we do not treat the dimension as a device to regulate
divergences: the beta functions obtained by our procedure are automatically finite.
Therefore, we can treat the dimension as an external continuous parameter
and ask questions about the structure of the FPs as this parameter
changes continuously.
Since we have the complete beta functions for any $D$, we could in principle
study the $D$-dependence without any approximation.
However, the beta functions for general $D$ are quite unwieldy and therefore we limit ourselves
to the first order of a Taylor expansion around four dimensions.
The main question we are interested in here is:
what becomes of the fixed points FP$_1$ and FP$_2$ when one moves away from
four dimensions?

To first order in $\epsilon$, the beta functions are
\bea
\beta_{\lambda} & =&  -\frac{133}{160\pi^2}\lambda^{2}
+\epsilon\,\Biggl[\lambda
+\frac{\lambda^2}{17280\pi^2\omega(1-2\omega)}
\Biggl(90
\nonumber\\
&&
+\omega\left(9094-7182\gamma+15475\log2+7182\log\pi-1111\log\frac{4(1+\omega)}{3}\right)
\nonumber\\
&&
-6\omega^2\left(3058-2394\gamma+4505\log2+2394\log\pi+283\log\frac{4(1+\omega)}{3}\right)
\nonumber\\
&&
-360\omega^3\left(1+2\log\frac{2(1+\omega)}{3}\right)
-80\omega^4\left(1+2\log\frac{2(1+\omega)}{3}\right)
\nonumber\\
&&
+60(1-2\omega)(1+70\omega)\theta
-6480\omega(1-2\omega)\theta^2
\Biggr)\Biggr]\ ,
\\
\beta_{\omega} & =&  -\frac{25 + 1098\,\omega+ 200\,\omega^2}{960\pi^2}\lambda
\nonumber\\
&&
+\epsilon\,
\frac{\lambda}{34560\pi^2\omega(1-2\omega)}
\Biggl[
\omega\left(1484-450\gamma+695\log2+450\log\pi+205\log\frac{4(1+\omega)}{3}\right)
\nonumber\\
&&
+4\omega^2\left(7055-4716\gamma+10310\log2+4716\log\pi-878\log\frac{4(1+\omega)}{3}\right)
\nonumber\\
&&
-12\omega^3\left(4618-2994\gamma+5525\log2+2994\log\pi+463\log\frac{4(1+\omega)}{3}\right)
\nonumber\\
&&
-80\omega^4\left(173-90\gamma+152\log2+90\log\pi+28\log\frac{4(1+\omega)}{3}\right)
\nonumber\\
&&
-160\omega^5\left(1+2\log\frac{2(1+\omega)}{3}\right)
\nonumber\\
&&
+60(1-2\omega)(5+22\omega+80\omega^2)\theta
-12960\omega^2(1-2\omega)\theta^2
\Biggr]\ ,
\\
\beta_{\theta} & = & \frac{7(56-171\,\theta)}{1440\pi^2}\lambda
+\epsilon\,
\frac{\lambda}{34560\pi^2\omega(1-2\omega)}
\Biggl[
\nonumber\\
&&
-\omega\left(8086-4704\gamma+10525\log2+4704\log\pi-1117\log\frac{4(1+\omega)}{3}\right)
\nonumber\\
&&
+6\omega^2\left(2662-1568\gamma+2855\log2+1568\log\pi+281\log\frac{4(1+\omega)}{3}\right)
\nonumber\\
&&
+360\omega^3\left(1+2\log\frac{2(1+\omega)}{3}\right)
+80\omega^4\left(1+2\log\frac{2(1+\omega)}{3}\right)
\nonumber\\
&&
+\Biggl(-60+2\omega\left(7954-7182\gamma+15475\log2+7182\log\pi-1111\log\frac{4(1+\omega)}{3}\right)
\nonumber\\
&&
-12\omega^2\left(2598-2394\gamma+4505\log2+2394\log\pi+283\log\frac{4(1+\omega)}{3}\right)
\nonumber\\
&&
-720\omega^3\left(1+2\log\frac{2(1+\omega)}{3}\right)
-160\omega^4\left(1+2\log\frac{2(1+\omega)}{3}\right)\Biggr)\theta
\nonumber\\
&&
+120(1-2\omega)(1+70\omega)\theta^2
-12960\omega(1-2\omega)\theta^3
\Biggr]\ .
\ena

All the terms in these beta functions contain at least one power of $\lambda$,
so $\lambda=0$ is a FP.
However, in this way the values of $\omega$ and $\theta$ remain undetermined.
In order to find FPs for the latter, we proceed as in four dimensions
and we use the rescaled parameter $\bt$.
Since $\beta_\omega$ and $\beta_\theta$ are linear in $\lambda$,
the equations $\frac{d\omega}{d\bt}=0$ and $\frac{d\theta}{d\bt}=0$
do not contain any power of $\lambda$ and can be solved numerically.
Since the terms of order $\epsilon$ contain $\log(1+\omega)$,
the equations are real only for $\omega>-1$.
The fixed point FP$_1$ has $\omega<-1$ and therefore can only
be real for $\epsilon$ strictly equal to zero.
We will not discuss this FP further.

\begin{figure}[htp]
\centering
\resizebox{0.9\textwidth}{!}{
\includegraphics{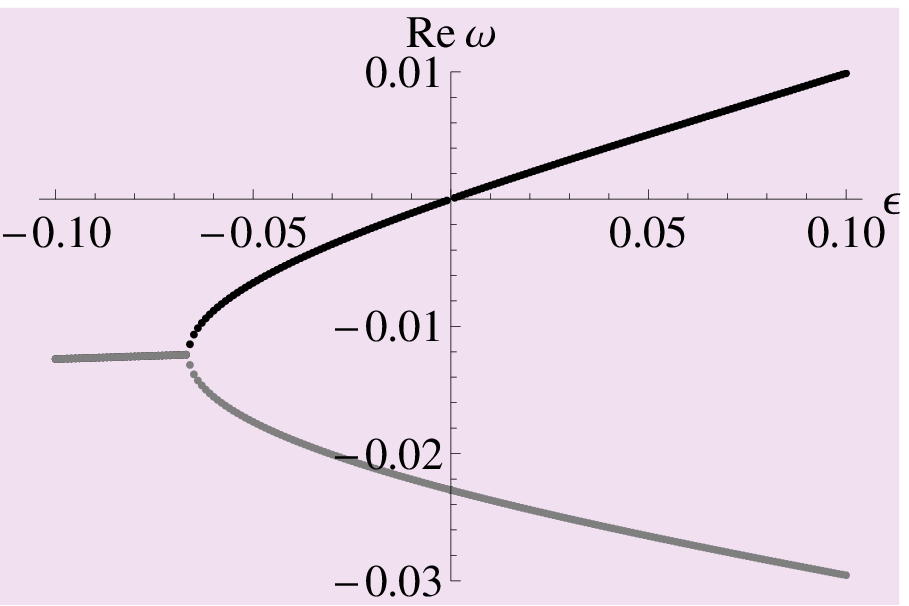}
\qquad\qquad
\includegraphics{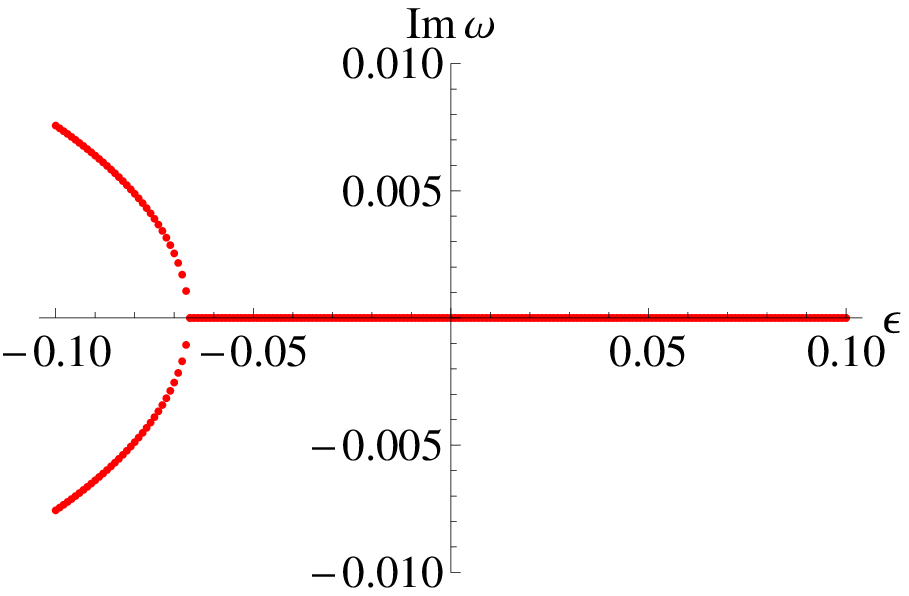}
}
\vskip1cm
\resizebox{0.9\textwidth}{!}{
\includegraphics{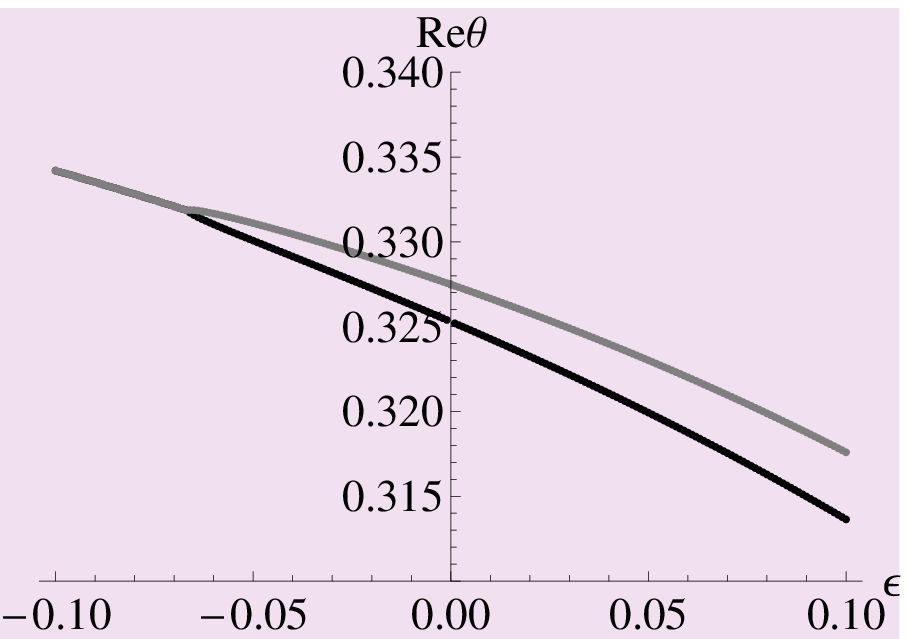}
\qquad\qquad
\includegraphics{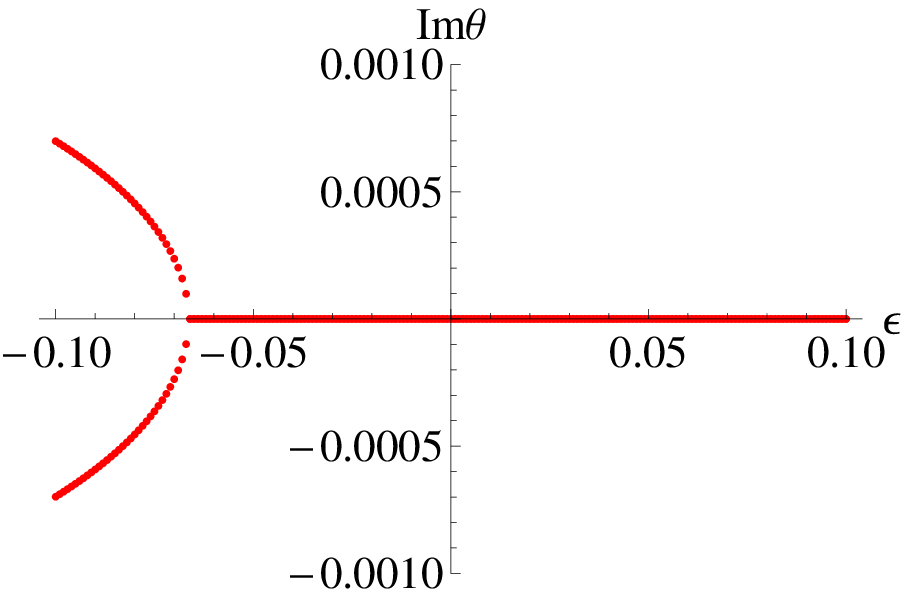}
}
\vskip1cm
\resizebox{0.9\textwidth}{!}{
\includegraphics{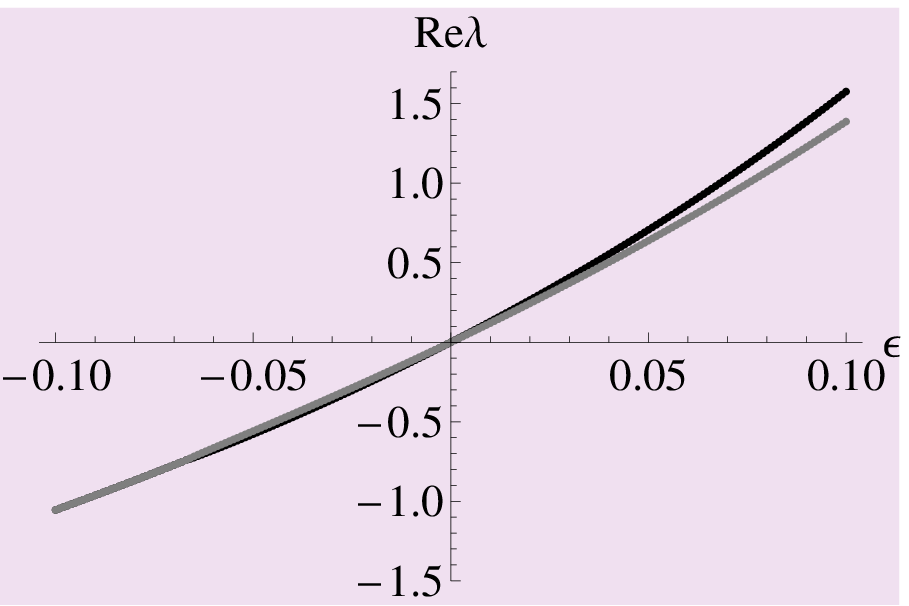}
\qquad\qquad
\includegraphics{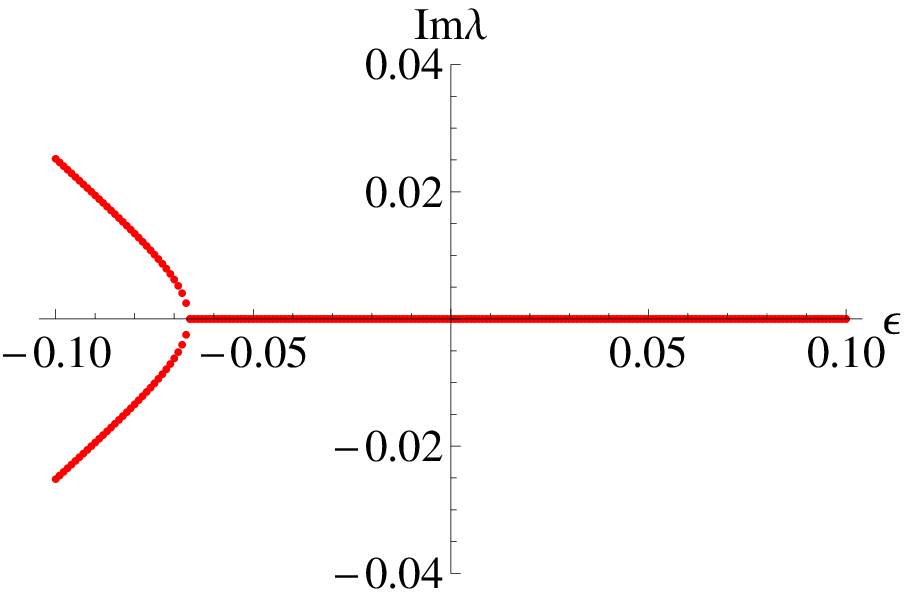}
}
\caption{The real and imaginary parts of $\omega_*$ (top),
$\theta_*$ (middle) and $\lambda_*$ (bottom) as functions of $\epsilon$.
Gray curves: FP$_2$, black curves: FP$_3$.}
\label{f:ripart}
\end{figure}

The fixed point FP$_2$ is real in a neighborhood of $D=4$,
however, for $\epsilon<-0.0663$ it becomes complex.
When one follows the evolution of this FP for increasing $\epsilon$,
one finds that the complex solution branches into two real solutions.
The more negative one is FP$_2$, the more positive one is a new FP
that we shall call FP$_3$.
When we consider the equation $\frac{d\lambda}{d\bt}=0$,
we find that all terms are linear in $\lambda$ except for the
first term in the square bracket, which is independent of $\lambda$.
Therefore, $\lambda=0$ is not a solution of $\frac{d\lambda}{d\bt}=0$.
There is however a nontrivial solution which can be found numerically
by using the FP values of $\omega$ and $\theta$.
The values of these FPs as functions of $\epsilon$ are shown in Fig.~\ref{f:ripart}.

\begin{figure}[htp]
\centering
\resizebox{0.9\textwidth}{!}{
\includegraphics{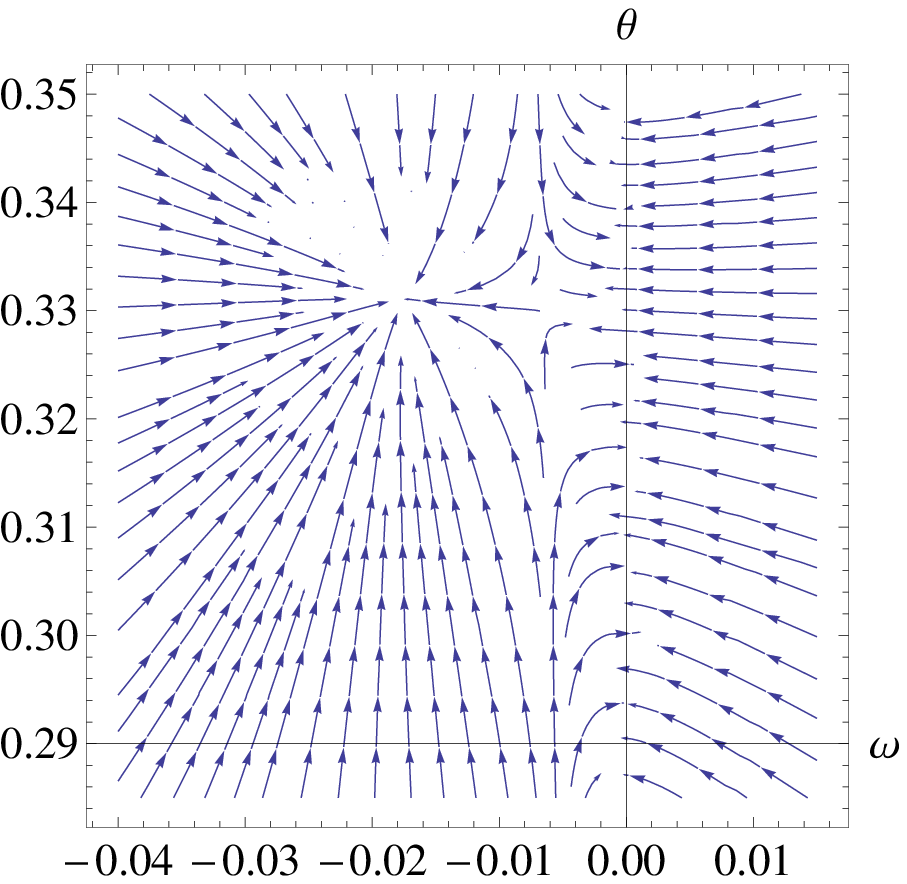}
\qquad\qquad
\includegraphics{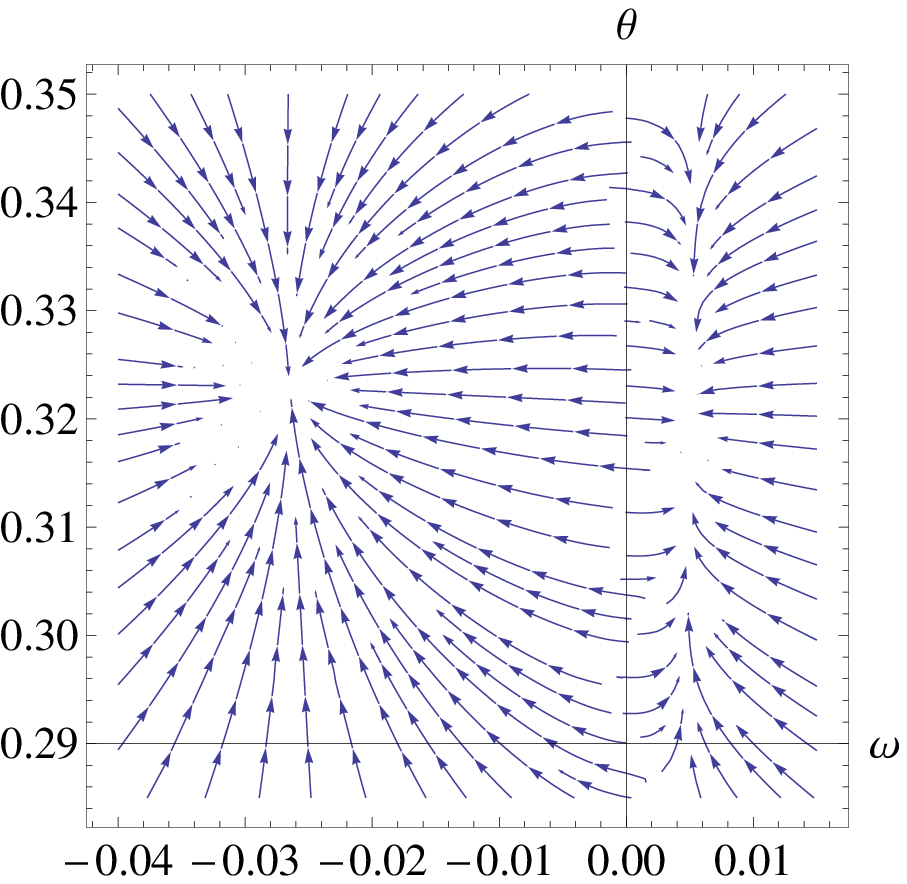}
}
\caption{The flow in the $\omega$-$\theta$ plane for $\epsilon=-0.05$ (left)
and $\epsilon=0.05$ (right). In both cases the line $\omega=0$ is a singularity
of the beta functions. Both FPs are UV-attractive for $\epsilon>0$;
FP$_2$ is UV attractive and FP$_3$ is mixed for $\epsilon<0$.
}
\label{f:epsilon}
\end{figure}

The analysis of the linearized flow near the FPs shows that
$\lambda$ is always an UV--attractive eigendirection.
For $\epsilon>0$, both FP$_2$ and FP$_3$ are UV--attractive.
(This can happen because they are separated by a repulsive singular line.)
For $\epsilon<0$, FP$_2$ is attractive and FP$_3$ is mixed,
with a relevant direction lying mostly in the $\lambda$-$\theta$ plane
and an irrelevant direction lying mostly in the $\lambda$-$\omega$ plane.
A picture of the flow in the $\omega$-$\theta$ plane is shown in Fig.~\ref{f:epsilon}.

The most interesting result of this analysis is that FP$_3$
has a natural continuation at $D=4$ where
$$
\mbox{FP}_3:(\lambda_*,\omega_*,\theta_*)=(0,0,0.325296)\ .
$$
This FP cannot be seen when one directly solves the equations in $D=4$.
The reason is that (in the absence of lower derivative terms, that do not affect the
evolution of the four--derivative terms anyway) $\omega=0$ corresponds to a Weyl--invariant
four--derivative sector.
Quantization of this theory requires that also Weyl--invariance be gauge fixed and gives rise
to different beta functions from the ones discussed in the preceding section.
When one continues to $D\not=4$, the theory is no longer Weyl--invariant and the general
quantization scheme discussed in this paper is valid.
This is a strong indication for the existence of a Weyl--invariant FP
in four dimensions. We plan to study this case separately in the future.


\section{Three dimensions}

Using that in three dimensions $\br_{\mu\nu\rho\sigma}^2=4\br_{\mu\nu}^2-\br^2$,
$\c$ automatically disappears from the expressions
of the beta functions when one makes the replacement $\a\to \a-\c, \b\to \b-4\c$.
We then obtain for $\tilde\alpha=k\alpha$ and $\tilde\beta=k\beta$:
\bea
\beta_{\tilde\alpha}
\hs{-2}&=&\hs{-2}
\tilde\alpha-
\frac{1}{960 \pi ^2}
\Bigg[
\left(1+\frac{47\tilde\alpha}{\tilde\beta}
+\frac{64\tilde\alpha^2}{\tilde\beta^2}
-\frac{128\tilde\alpha}{2 \tilde\a+\tilde\beta}
\right)\sqrt{-\frac{2\tilde\beta}{\tilde\alpha}}
\nn
&&
+8\Bigg( 129+\frac{46\tilde\alpha}{\tilde\beta}
-\frac{114\tilde\alpha^2}{\tilde\beta^2}
+\frac{32\tilde\a}{2\tilde\a+\tilde\b}
+\tilde\alpha\frac{131\tilde\alpha+49\tilde\beta}{(8\tilde\alpha+3\tilde\beta)^2}
\Bigg)\Bigg],
\ena
\bea
\beta_{\tilde\beta}
\hs{-2}&=&\hs{-2}
\tilde\beta+
\frac{1}{1440\pi^2}
\Bigg[
-3\left(1
-\frac{113\tilde\a}{\tilde\b}
+\frac{64\tilde\a^2}{\tilde\b^2}
+\frac{192\tilde\a}{2\tilde\a+\tilde\b}
\right) \sqrt{-\frac{2\tilde\beta}{\tilde\alpha}}
\nn
&&
+8\Bigg(
413 - \frac{93\tilde\a}{\tilde\b}
+\frac{72\tilde\a^2}{\tilde\b^2}
+\frac{144\tilde\a}{2\tilde\alpha+\tilde\beta}
+\tilde\a\frac{652\tilde\alpha+243\tilde\beta}{(8\tilde\alpha+3\tilde\beta)^2}\Bigg)
\Bigg].
\ena
These form a closed system of equations with a FP at
$\tilde\alpha=0.0496$, $\tilde\beta=-0.1381$.

The beta functions of $\tilde\Lambda\equiv\Lambda/k^2$ and
$\tilde G\equiv G\,k$ are
\bea
\beta_{\tilde \Lambda} & = &
-2\tilde\Lambda
+p(\tilde\alpha,\tilde\beta)\tilde\Lambda
-q(\tilde\alpha,\tilde\beta)\tilde G \tilde\Lambda
+r(\tilde\alpha,\tilde\beta)\tilde G
+s(\tilde\alpha,\tilde\beta)
+\frac{t(\tilde\alpha,\tilde\beta)}{\tilde G}\ ,
\\
\beta_{\tilde G} & = & \tilde G-u(\tilde\alpha,\tilde\beta)\tilde G
-q(\tilde\alpha,\tilde\beta)\tilde G^2
\ ,
\ena
where
\bea
p(\tilde\alpha,\tilde\beta)
\hs{-2}&=&\hs{-2}
\frac{-2304\tilde\alpha^3+1120\tilde\alpha^2 \tilde\beta
+1778\tilde\alpha\tilde\beta^2 +387\tilde\beta^3
}{48\pi^2(8\tilde\a+3\tilde\b)^2 \tilde\beta^2},
\\
q(\tilde\alpha,\tilde\beta)
\hs{-2}&=&\hs{-2}
\frac{1}{3\pi\s}
\left[37 + \frac{\tilde\b}{8\tilde\a+3\tilde\b}
-\frac{16\tilde\a}{\tilde\b}
+\sqrt{-\frac{2\tilde\b}{\tilde\a}}
+4 \left(1-\frac{8\tilde\a}{\tilde\b} \right)\sqrt{\frac{-2\tilde\a}{\tilde\b}}
\right],~~~~
\\
r(\tilde\alpha,\tilde\beta)
\hs{-2}&=&\hs{-2}
\frac{1}{3\pi^2}\left[24-\frac12 \left(\frac{-2\tilde\b}{\tilde\a}\right)^{3/2}
-8 \left(\frac{-2\tilde\a}{\tilde\b}\right)^{3/2} \right],
\\
s(\tilde\alpha,\tilde\beta)
\hs{-2}&=&\hs{-2}
-\s\frac{16\tilde\alpha+5\tilde\beta}{8\pi^2(8\tilde\alpha+3\tilde\beta)\tilde\beta},
\\
t(\tilde\alpha,\tilde\beta)
\hs{-2}&=&\hs{-2}
3\s^2\frac{128\tilde\alpha^2+96\tilde\alpha\tilde\beta+19\tilde\beta^2}
{1024\pi^3(8\tilde\alpha+3\tilde\beta)^2 \tilde\beta^2},\\
u(\tilde\alpha,\tilde\beta)
\hs{-2}&=&\hs{-2}
\frac{2304\tilde\alpha^3 +2912\tilde\alpha^2\tilde\beta
+1174\tilde\alpha\tilde\beta^2 +153\tilde\beta^3}
{48\pi^2(8\tilde\alpha+3\tilde\beta)^2 \tilde\beta^2}.
\ena

\begin{figure}[t]
\begin{center}
\includegraphics[width=6cm]{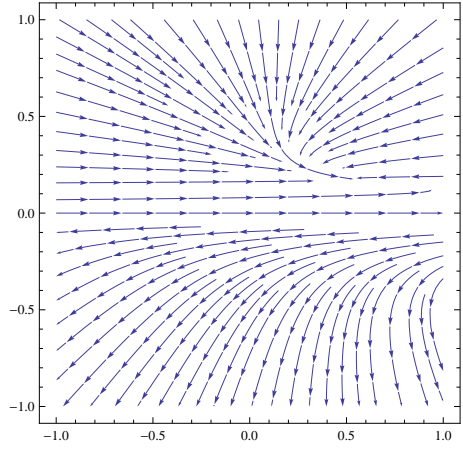}
\caption{The flow in the $(\tilde\Lambda,\tilde G)$--plane
when $\tilde\alpha$ and $\tilde\beta$ are at FP$_1$}
\label{fig2}
\end{center}
\end{figure}

If we choose $\tilde\alpha$ and $\tilde\beta$ at their FP,
one finds a FP at $\tilde\Lambda=0.5355\s$ and $\tilde G=0.1758\s$,
with eigenvalues $-2.866$ and $-1.235$. Thus this is a UV--stable FP.
This flow is depicted in Fig.~\ref{fig2}.
%
Note that there is no Gaussian FP in this plane because $\tilde\alpha$ and $\tilde\beta$
are already at a nontrivial FP.

Let us change coordinates and consider the beta functions of the couplings
$\lambda_3=1/\b$ and $\omega_3=-2\a/\b$.
These coupling constants are similar but different from those defined in the action~\p{action1}.
We find that they are given by
\bea
\b_{\la_3}
\hs{-2}&=&\hs{-2}
-\la_3 - \frac{\la_3^2}{480\pi^2 \sqrt{\omega_3}(\omega_3-1)}
\Big[ 2-81\omega_3-81\omega_3^2-32\omega_3^3 \nn
&&\hs{-2} +\frac{8\sqrt{\omega_3}(-1239+4660\omega_3-5720\omega_3^2+2219\omega_3^3
+8\omega_3^4+96\omega_3^5)}{(4\omega_3-3)^2} \Big], \\
\b_{\omega_3}
\hs{-2}&=&\hs{-2}
-\frac{\la_3}{480\pi^2 \sqrt{\omega_3}(\omega_3-1)}
\Big[ (1+\omega_3)(2+79\omega_3-81\omega_3^2-32\omega_3^3) \nn
&&\hs{-2} +\frac{2\sqrt{\omega_3}(4644-23486\omega_3+43055\omega_3^2-32209\omega_3^3+5788\omega_3^4
+1856\omega_3^5+384\omega_3^6)}{(4\omega_3-3)^2} \Big].~~
\ena
We see that the beta functions vanish for $\la_3=0$, leaving $\omega_3$ undetermined.
In this variable, we see that we can have the Gaussian FP for $\la_3=0$ with fixed $\omega_3$,
corresponding to $\tilde\b\to \infty$ with $\tilde\a/\tilde\b$ = fixed and $s=t=0$.
As in four dimensions, using the variable $\bt$ defined by $d\bt=\lambda_3 dk/k$
removes one power of $\lambda_3$ from the beta functions and
one finds a FP at $(\la_3,\omega_3)=(-7.240, 0.7187)$.
Another important result is that $8\tilde\a+3\tilde\b$ does not vanish at the FP.
This implies that the new massive gravity does not correspond to the FP.
This was suggested in \cite{Ohta3}, and we confirm it.

\section{Five dimensions}

In dimensions $D\not= 4$, $\omega$ and $\theta$ remain dimensionless but
$\lambda$ has dimension $4-D$, so we define $\tilde\lambda=\lambda\,k^{D-4}$.
We find the following beta functions:
\bea
\beta_{\tilde\lambda} &=& \tilde\lambda+\frac{\tilde\lambda ^2}{3360(4 \pi)^3 \omega ^2(\omega -1)}
\Big[  -60 (\omega-1)(621 \omega^2 - 42 \omega +1) \theta^2 \nn
&& + 80 ( 279 \omega-2)(\omega -1)  \omega\theta
-(270 \omega^4 + 1290 \omega^3 +131258 \omega^2 -112798 \omega +460)\omega
\nn
&&
+\sqrt{2} \omega ^2 \frac{243 \omega^4
+1872 \omega^3
+8922 \omega^2
+18304 \omega
+11619 }{\sqrt{3\omega+5}}
\Big], \\
\b_\omega &=&
\frac{\tilde\lambda}{3360(4 \pi)^3(\omega -1) \omega ^2}
\Big[ 60(1-\omega)(621 \omega^3 +147 \omega^2-17 \omega +9)\t^2 \nn
&& +40(1-\omega)( 9\omega^2 -446 \omega-63) \omega\t \nn
&&
-(270 \omega^4 +30810 \omega^3+130058 \omega^2-124370 \omega-4000)\omega^2
 \nn
&&
+\sqrt{2} \omega^2 \frac{243 \omega^5 +2925 \omega^4+15690 \omega^3+32326 \omega^2+20003 \omega
-5651}{\sqrt{3 \omega +5}} \Big], \nn
\b_\t
&=&
\frac{\tilde\lambda}{20160 (4\pi)^3 (\omega -1) \omega ^2}
\Big[ -360 (\omega-1) (621 \omega^2-42 \omega +1)\theta ^3 \nn
&&
-12 (135 \omega^4 +645 \omega^3 +70189 \omega^2 -60679 \omega-50) \t  \omega \nn
&&
+ 60 (\omega-1)(2799 \omega^2-106 \omega+3) \t^2
+ 10 (81 \omega^3 +387 \omega^2 +23103 \omega -17427) \omega^2 \nn
&&
+6 \sqrt{\frac{2}{3\omega+5}} \Big( 243 \omega^4 +1872 \omega^3
+8922 \omega^2 +18304 \omega +11619 \Big) \t \omega^2 \nn
&&
-3 \sqrt{\frac{2}{3\omega+5}} ( 243 \omega^4 +1872 \omega^3 +8698 \omega^2
+17856 \omega +12291) \omega ^2 \Big].
\ena
In the standard RG time $t=\log k$, these beta functions
have a FP for $\tilde\lambda=0$ and undetermined values of $\omega$ and $\theta$.
The flow in this sector is therefore studied more conveniently
using the variable $\bt$, which has the effect of removing one power of $\tilde\lambda$
from the beta functions.
There is then no FP with $\tilde\lambda=0$ but there are four FPs
with coordinates given in the following table:
\begin{center}
    \begin{tabular}{ | l | r | r | r |}
    \hline
     & $\tilde\lambda_*$\hs{5} & $\omega_*$\hs{3} & $\theta_*$\hs{3} \\
      \hline
    FP$_1$ & $1.2654$   & $1007.9$ & $-406.9$ \\ \hline
    FP$_2$ & $-0.61121$ & $1796.0$ & $841.4$ \\ \hline
    FP$_3$ & $-9.4483$  & $27.9$ & $0.54$ \\ \hline
    FP$_4$ & $53.53$    & $-0.082$ & $0.27$ \\ \hline
    \end{tabular}
\end{center}

For $\tilde\Lambda$ and $\tilde G$ we have the beta functions:
\bea
\b_{\tilde\Lambda}
&=&
-2 \tilde\Lambda
-\frac{\tilde\lambda  \sigma  (9 \omega +1)}{64 \pi^3 \omega}
+\frac{15 \tilde\lambda^2 \sigma^2 (9 \omega^2+1)}{32768 \pi^4 \tilde G \omega^2}
+\frac{\tilde\Lambda  \Big[ 6(9 \omega^2+1) \t + \omega( 135\omega^2+280 \omega+3) \Big]}
{1024 \pi^3 \omega^2} \tilde\lambda
\nn &&
-\tilde G \frac{ \sqrt{2} (3\omega +5)^5 -2304(3 \omega +5)^{5/2} +16384\sqrt{2}}
{960 \pi^2(3 \omega +5)^{5/2}}
-\frac{\tilde G \tilde\Lambda}{1440 \pi^2 \sigma  \omega}
\Big[144 \theta (9 \omega +1) \nn
&&
+ (2952 \omega+3240)\omega
+\sqrt{\frac{2}{3\omega+5}}(81 \omega^3+ 495 \omega^2 +975 \omega+625)\omega
+ \frac{2560\sqrt{2}\omega}{(3\omega+5)^{3/2}}
\Big] ,~~~~ \\
\b_{\tilde G}
&=&
3\tilde G + \frac{\tilde G}{1024 \pi^3 \omega^2}
\Big(6 \theta (9 \omega^2+1)
+ \omega (135 \omega^2-180 \omega -17) \Big)\tilde\lambda
-\frac{\tilde G^2}{1440 \pi^2 \sigma \omega} \Big[144 \theta (9 \omega +1) \nn
&&
+\omega (2952 \omega+3240) +\sqrt{\frac{2}{3\omega+5}}(81 \omega^3
+495 \omega^2+975\omega+625)\omega + \frac{2560\sqrt{2}\omega}{(3\omega+5)^{3/2}}
 \Big].
\ena
The Gaussian FP corresponds to $\tilde\lambda=\tilde G=\tilde\Lambda=0$,
with arbitrary values of $\omega$ and $\theta$.
One finds that the eigenvalues of the linearized flow equation at this point
are equal to $3$, $-2$, $1$, $0$, $0$, which are just the opposites of the
canonical dimensions of $G$, $\Lambda$, $\lambda$, $\omega$ and $\theta$,
as expected.

Next we consider FPs with $\tilde\lambda_*\not=0$.
We take the FP values of $\tilde\lambda$, $\omega$ and $\theta$
given in the table above and use them in the beta functions of $\tilde G$, $\tilde\Lambda$.
One finds various FPs with $\tilde G$ and $\tilde\Lambda$
either zero or nonzero.
If one chooses $\tilde G$ and $\tilde\Lambda$ equal to zero,
the matrix of the derivatives of the beta functions has
singularities in some off-diagonal elements
(more precisely the derivatives of $\tilde\Lambda$ with respect to $\tilde\lambda$,
$\omega$ and $\tilde G$).
We will not consider this case further.
The FPs with non-zero $\tilde\Lambda$ and $\tilde G$ are
\begin{center}
    \begin{tabular}{ | l | r | r | r |}
    \hline
      & $\tilde G_*$~~~~~~~ & $\tilde\Lambda_*$~~~ \\
      \hline
    FP$_1$ & $4.992\times 10^{-5}$ & $-0.4839$  \\ \hline
    FP$_2$ & $-3.976\times10^{-6}$ & $-0.5397$  \\ \hline
    FP$_3$ & $6.619\times10^{-2}$ & $-0.1201$ \\ \hline
    FP$_4$ & $13.74$ & $0.6038$ \\ \hline
    \end{tabular}
\end{center}
(These are the values for $\sigma=1$; for $\sigma=-1$ all the
elements of the table have opposite sign.)

Of these FPs, the second and the third are physically uninteresting
because of the sign of $\tilde\lambda_*$. The first three also have suspiciously
small values for $\tilde G$ and the first two also have one very large critical
exponent (of order $10^3$). It is possible that these are spurious FPs of the
type that often appear in polynomial truncations.
The fourth is more promising: the values of $\tilde G$ and $\tilde\Lambda$
are closer to those found in the Einstein-Hilbert truncation.
The critical exponents at this FP are as follows:
the cosmological constant is an eigendirection with eigenvalue $-4.635$;
the remaining eigenvalues are $-3.460$ in the direction of $\tilde G$, with a small
mixture of $\tilde\Lambda$; $-3.150$ in the direction of $\tilde G$, with a small
mixture of all the other operators;
$-1$ in the direction of $\tilde\lambda$ with a small mixture of $\tilde\Lambda$ and $\tilde G$;
$-0.9870$ in the direction of $\tilde\lambda$ with small mixtures of all the other operators.
This FP is therefore UV attractive in all five directions considered.

\section{Six Dimensions}

In dimension $D=6$, we have
\bea
\b_{\tilde\lambda}
&=&
2\tilde\lambda+
\frac{\tilde\lambda ^2}{243\times 10^3(4 \pi)^3 \omega ^2}
\Big[-1215 \theta ^2(24 \omega -1) (32 \omega -3)+2700 \theta  \omega (212 \omega -3)
\nn &&
+\frac{\omega}{4\omega+9} (2048 \omega^5 +20352 \omega^4 +181728 \omega^3 -7130808 \omega^2
-16720182 \omega +54675) \Big], \\
\beta_\omega
&=&
-\frac{\tilde\lambda}{243\times 10^3(4 \pi)^3 \omega ^2}
\Big[3645 \theta ^2 (\omega +1)\{16 \omega  (16 \omega -3)+21\}
+2700 \theta  \omega \{\omega  (124 \omega -387)-36\} \nn
&&
-\frac{\omega^2}{4\omega+9} (2048 \omega^5 +32640 \omega^4 -1510560 \omega^3
-10360440 \omega^2-15408630 \omega +7533) \Big], \\
\b_\t
&=&
\frac{\tilde\lambda}{486\times 10^3(4 \pi)^3 \omega ^2}
 \Big[-2430 \theta ^3 (24 \omega -1) (32 \omega -3)
+135 \theta ^2 \{4 \omega (2918 \omega -219)+27 \} \nn
&&
-32 \omega ^2 (16 \omega^3 +123\omega^2+1098 \omega -28134) \nn
&&
+\frac{2 \theta  \omega}{4\omega+9}
(2048 \omega^5 +20352 \omega^4 +181728 \omega^3 -7919208 \omega^2
-18526482 \omega -18225)
\Big].
\ena
Proceeding as in the five--dimensional cases one finds seven FPs:
\begin{center}
    \begin{tabular}{ | l | r | r | r |}
    \hline
     & $\tilde\lambda_*$\hs{5} & $\omega_*$\hs{3} & $\theta_*$\hs{3} \\
      \hline
    FP$_1$ & $148.358$ & $105.0$ & $-25.98$ \\ \hline
    FP$_2$ & $-37.24$ & $265.4$ & $103.0$ \\ \hline
    FP$_3$ & $55.54$ & $-33.17$ & $0.4701$\\ \hline
    FP$_4$ & $-106.647$ & $24.58$ & $0.5544$\\ \hline
    FP$_5$ & $502.2$ & $-4.545$ & $0.2493$\\ \hline
    FP$_6$ & $396.5$ & $-2.268$ & $0.1871$\\ \hline
    FP$_7$ & $513.725$ & $-0.1376$ & $0.2448$\\ \hline
    \end{tabular}
\end{center}

Upon substituting these values in the beta functions for $\tilde\Lambda$ and $\tilde G$.
\bea
\b_{\tilde\Lambda}
\hs{-2}&=&\hs{-2}
-2 \tilde\Lambda
-\frac{\tilde\lambda  \sigma  (28 \omega +3)}{768 \pi^3 \omega}
+\frac{\tilde\lambda ^2 \sigma ^2 (56 \omega^2+9)}{49152 \pi^4 \tilde G \omega^2}
+\frac{\tilde\Lambda  \Big[3(56 \omega^2+9)\t +(168 \omega^2+286 \omega+3) \omega \Big]}
{5760 \pi^3 \omega^2} \tilde\lambda \nn
&&
+\frac{\tilde G}{\pi^2}\Big[ \frac{9}{16}-\frac{1125}{16(4\omega +9)^3}
-\frac{(4\omega+9)^3}{81000}\Big]
-\frac{\tilde G\tilde\Lambda}{81000\pi^2\s\omega}\Big[
1350 \theta (28 \omega +3)
\nn &&
+\frac{\omega}{ (4 \omega +9)^2} (2048 \omega^5+26880 \omega^4 + 714240 \omega^3+3346920 \omega^2
+5175900 \omega + 2646513)\Big] ,~~~~~~ \\
\b_{\tilde G}
\hs{-2}&=&\hs{-2}
4\tilde G+ \frac{\tilde G}{5760 \pi^3 \omega^2}
\Big[3 \theta (56 \omega^2+9) +2 \omega (84 \omega^2-163 \omega -12) \Big]\tilde\lambda
-\frac{\tilde G^2}{81000 \pi^2 \sigma  \omega}
\Big[1350 \theta  (28 \omega +3) \nn
&&
+\frac{\omega}{(4\omega+9)^2} (2048 \omega^5+26880 \omega^4 + 714240 \omega^3+3346920 \omega^2
+5175900 \omega + 2646513)\Big].
\ena
one finds the following values of $\tilde G_*$ and $\tilde\Lambda_*$:
\begin{center}
    \begin{tabular}{ | l | r | r | r
     |}
    \hline
      & $\tilde G_*$\hs{7} & $\tilde\Lambda_*$\hs{6} \\
      \hline
    FP$_1$ & $0.0722003$ & $-0.679014$  \\ \hline
    FP$_2$ & $-0.00284264$ & $-0.211878$  \\ \hline
    FP$_3$ & $-0.372332$ & $-0.183136$ \\ \hline
    FP$_4$ & $0.374408$ & $-0.0119273$ \\ \hline
    FP$_5$ & $-6.84831$ & $-0.342368$ \\ \hline
    FP$_6$ & $0.0269567$ & $129.099$ \\ \hline
    FP$_7$ & $123.127$ & $1.16044$ \\ \hline
    \end{tabular}
\end{center}
(These are the values for $\sigma=1$; for $\sigma=-1$ all the
elements of the table have opposite sign.)

Of these fixed points, FP$_3$, FP$_4$, FP$_6$ and FP$_7$ are UV attractive in all directions;
the others have one or two irrelevant directions.

\section{Conclusions}

We have studied the one-loop beta functions in higher derivative gravity for general
background in arbitrary dimensions.
The results for $D=4$ agree with the literature, up to minor differences due to the
scheme--dependence of certain coefficients.
In the four--derivative sector, there are at one loop two well--known
fixed points FP$_1$ and FP$_2$.
To each of these there correspond two FPs in the $\tilde\Lambda$--$\tilde G$ planes.
There is a Gaussian FP, which is UV--attractive in the $\tilde\Lambda$ direction
and repulsive along a complementary direction, making $\tilde G$
(with a small mixture of $\tilde\Lambda$) an irrelevant coupling.
There is also a non-Gaussian FP that is UV--attractive in both directions.
Altogether the qualitative structure of the flow in the $\tilde\Lambda$--$\tilde G$ planes
is surprisingly similar to that of the theory in the Einstein--Hilbert truncation.
The main difference is the reality of the critical exponents,
to be contrasted with the complex conjugate critical exponents seen in the
Einstein--Hilbert truncation of the flow,
which lead to spiralling trajectories near the non-Gaussian FP.
It may be that the approximation used here is too crude.
On the other hand, recent studies of the Einstein--Hilbert truncation
taking into account the anomalous dimension of the fluctuation field $h_{\mu\nu}$
find real critical exponents in that case too \cite{clpr,cdp}.
(See also the general argument in \cite{bene}.)
To settle the question of the correct critical exponents, one may have to
perform more complicated calculations, for example
the full truncated RG of higher derivative gravity taking
into account the anomalous dimension of the fluctuation field.
Another issue that could be clarified by such a calculation is the position of the
FP of the dimensionless coupling $\tilde\lambda$, which was found to occur
at some non-zero value in \cite{bms}.

In this paper we have extended the calculation of one-loop beta functions of
higher derivative gravity to arbitrary dimension.
Insofar as the beta functions can be obtained directly as finite quantities,
without having to refer to divergences and regulators,
the dimension can be treated formally as a continuous parameter of the theory.
In this way the structure of the FPs can be reliably studied as a function of dimension.
There has been one previous study along these lines \cite{BS2}, but our results
differ significantly.
This can be traced to the structure of the heat kernel coefficients
of fourth--order and of non--minimal second order operators, that we have
taken from \cite{Gusynin1990,GK1999}.
We find that FP$_1$ becomes complex as soon as one departs from $D=4$,
but FP$_2$ remains real in a neighborhood of $D=4$.
Furthermore, for arbitrarily small $\epsilon$, there is another fixed point FP$_3$
which does not solve our beta functions in $D=4$.
We believe that this FP must correspond to a Weyl--invariant theory,
which requires a different quantization procedure.
We hope to return on this point in the future.

One of the main motivations of this work was the structure of higher derivative
gravity in $D=3$. Due to the fact that in $D=3$, $E=0$ identically, there are
only two independent beta functions in the four--derivative sector,
and the FP structure is different from $D=4$.
In fact, when $D\lesssim 3.93$ the fixed points FP$_2$ and FP$_3$
merge and become complex.
Nevertheless, there is also in $D=3$ a non-Gaussian FP with
non--vanishing values of $\tilde\alpha$ and $\tilde\beta$ (recall that the
four--derivative couplings are dimensionful in $D=3$).
The corresponding flow in the $\tilde\Lambda$--$\tilde G$ planes
has a non--Gaussian FP that is somewhat similar to the
one that is found in the Einstein--Hilbert truncation or
in the Chern--Simons theory, except that it has a relatively large
value of $\tilde\Lambda$.
An important point is that new massive gravity does not
correspond to a FP, confirming the result in \cite{Ohta3}.

On the other hand, when one goes to dimensions greater than four
the fixed points FP$_2$ and FP$_3$ remain real but there appear
further FPs.
There is always a Gaussian FP with $\tilde\lambda=\tilde\Lambda=\tilde G=0$
and unspecified $\omega$ and $\theta$, but there are also FPs
with $\tilde\lambda\not=0$.
It is difficult, within the present analysis, to establish which ones of these are physical
and which ones are mere truncation artifacts.

\section*{Acknowledgment}

We would like to thank Ilya Shapiro for valuable discussions.
Part of this work was carried out while the authors were attending
the workshops ``Topics in Holography, Supersymmetry and Higher Derivatives''
at Texas A\&M University, and ``2nd Mediterranean Conference on Classical and Quantum Gravity''
at Veli Lo\v{s}inj, Croatia.
We thank the organizers for invitation and providing stimulating atmosphere.
This work was supported in part by the Grant-in-Aid for
Scientific Research Fund of the JSPS (C) No. 24540290, and (A) No. 22244030.

\appendix

\section{Conventions and useful formulae}
\label{conv}

Here we summarize our conventions and formulae necessary in the text.
We give these such that they are valid for any dimension $D$.

Our signature of the metric is $(-,+,\dots +)$ and the curvature tensors are given as
\bea
R^\a{}_{\b\mu\nu} &=&
\pa_\mu \G^{\a}_{\b\nu} - \pa_\nu \G^{\a}_{\b\mu}
+ \G^{\a}_{\mu\la} \G^{\la}_{\b\nu} - \G^{\a}_{\nu\la} \G^{\la}_{\b\mu}, \nn
R_{\mu\nu} &=& R^\a{}_{\mu\a\nu}.
\ena
The backgrounds are denoted with overbar.
Expansion around the background gives
\bea
\G^\a_{\mu\nu}
&=& \bar \G^\a_{\mu\nu} + \G^{\a (1)}_{\mu\nu} + \G^{\a(2)}_{\mu\nu},
\ena
where
\bea
\G^{\a (1)}_{\mu\nu} &=& \frac12 (\nabla_\nu h^\a{}_\mu
 +\nabla_\mu h^\a{}_\nu-\nabla^\a h_{\mu\nu}), \\
\G^{\a(2)}_{\mu\nu} &=& -\frac12 h^{\a\b} (\nabla_\nu h_{\mu\b}
+\nabla_\mu h_{\nu\b}-\nabla_\b h_{\mu\nu}).
\ena
Note that
\bea
\sqrt{-g} = \sqrt{-\bg} \Big[ 1+\frac{1}{2}h
+ \frac{1}{8}(h^2-2 h_{\mu\nu}^2) + O(h^3) \Big].
\ena
We find, to the second order,
\bea
R^\mu{}_{\nu\a\b}\!\! &=&\!\! \br^\mu{}_{\nu\a\b} + R^\mu{}_{\nu\a\b}^{(1)}
+ R^\mu{}_{\nu\a\b}^{(2)}, \nn
R^\mu{}_{\nu\a\b}^{(1)}\!\! &=&\!\! \frac12 (\nabla_\a\nabla_\nu h^\mu_\b
- \nabla_\a\nabla^\mu h_{\nu\b} - \nabla_\b \nabla_\nu h^\mu_\a
+ \nabla_\b\nabla^\mu h_{\nu\a}) +\frac12 \br_{\nu\c\a\b} h^{\mu\c}
+\frac12 \br^\mu{}_{\c\a\b} h^\c_\nu,~~~ \\
R^\mu{}_{\nu\a\b}^{(2)}\!\! &=&\!\!
\nabla_\a \G^{\mu(2)}_{\nu\b} - \nabla_\b \G^{\mu(2)}_{\nu\a}
+ \G^{\mu(1)}_{\la\a} \G^{\la(1)}_{\nu\b}
- \G^{\mu(1)}_{\la\b} \G^{\la(1)}_{\nu\a} \nn
&=&\!\! -\frac12 h^{\mu\c}\nabla_\a (\nabla_\b h_{\nu\c}+\nabla_\nu h_{\b\c}
-\nabla_\c h_{\nu\b})
-\frac14 \nabla_\a h^{\mu\c}(\nabla_\b h_{\nu\c}+\nabla_\nu h_{\b\c}
-\nabla_\c h_{\nu\b}) \nn
&&\hs{-2} +\; \frac14 \nabla_\c h^\mu_\a (\nabla_\b h^\c_\nu+\nabla_\nu h^\c_\b
-\nabla^\c h_{\nu\b})
-\frac14 \nabla^\mu h_{\a\c} (\nabla_\b h^\c_\nu+\nabla_\nu h^\c_\b
-\nabla^\c h_{\nu\b}) \nn
&& - (\a \leftrightarrow \b).
\ena
Though the last one looks different from that in \cite{BS1}, it coincides.
Similarly
\bea
R^{(1)}_{\mu\nu}\!\! &=&\!\! -\frac12 (\nabla_\mu \nabla_\nu h
- \nabla_\mu h_{\nu} - \nabla_\nu h_\mu + \Box h_{\mu\nu})
- \br_{\a\mu\b\nu}h^{\a\b}+\frac12 \br_{\mu\a}h^\a_\nu +\frac12 \br_{\nu\a} h^\a_\mu
, \nn
R^{(2)}_{\mu\nu}\!\! &=&\!\! \frac12 \nabla_\mu(h^{\a\b} \nabla_\nu h_{\a\b})
-\frac12 \nabla_\a \{h^{\a\b}( \nabla_\mu h_{\nu\b}+ \nabla_\nu h_{\mu\b}
- \nabla_\b h_{\mu\nu}) \} \nn
&& \hs{-10}
- \frac14 (\nabla_\mu h^\b_\a+ \nabla_\a h^\b_\mu -\nabla^\b h_{\a\mu})
(\nabla_\b h^\a_\nu + \nabla_\nu h^\a_\b -\nabla^\a h_{\b\nu})
+ \frac14 \nabla_\a h (\nabla_\mu h^\a_\nu + \nabla_\nu h_\mu^\a -\nabla^\a h_{\mu\nu}), \nn
R^{(1)} \!\! &=&\!\! \nabla_\mu h^\mu -\Box h - \br_{\mu\nu} h^{\mu\nu}, \nn
R^{(2)} \!\! &=&\!\! \frac12 \nabla_\mu(h^{\a\b} \nabla^\mu h_{\a\b})
-\frac12 \nabla_\a \{h^{\a\b}( 2 h_\b - \nabla_\b h) \}
- \frac14 (\nabla_\mu h^\b_\a+ \nabla_\a h^\b_\mu -\nabla^\b h_{\a\mu}) \nabla_\b h^{\a\mu}
 \nn
&& \hs{-10}
+\; \frac14 ( 2 h^\a -\nabla^\a h)\nabla_\a h
+\frac12 h^{\a\b}\nabla_\a \nabla_\b h
-\frac12 h_\a^\mu \nabla_\b ( \nabla^\a h_\mu^\b +\nabla_\mu h^{\a\b}
-\nabla^\b h^\a_\mu) + \br_{\mu\nu} h^\mu_\a h^{\nu\a} \nn
&=& \hs{-2}
\frac34 \nabla_\a h_{\mu\nu} \nabla^\a h^{\mu\nu} +h_{\mu\nu} \Box h^{\mu\nu}
-h_\mu^2 + h_\mu \nabla^\mu h -2 h_{\mu\nu} \nabla^\mu h^\nu
+h_{\mu\nu} \nabla^\mu \nabla^\nu h \nn
&& \hs{-2}
-\; \frac12 \nabla_\mu h_{\nu\a} \nabla^\a h^{\mu\nu} -\frac14 \nabla_\mu h \nabla^\mu h
+\br_{\a\b\c\d} h^{\a\c} h^{\b\d}.
\ena
where $\Box \equiv \nabla_\mu \nabla^\mu$.
Note that $\bg^{\mu\nu} R^{(1)}_{\mu\nu} \neq R^{(1)}$,
because the latter has additional contribution from $h^{\mu\nu} \br_{\mu\nu}$.
When total derivative terms are dropped, $R^{(2)}$ makes the contribution to the action
\bea
R^{(2)} \simeq
\frac14 ( h_{\mu\nu}\Box h^{\mu\nu} +h \Box h + 2 h_\mu^2
+2 \br_{\a\b}h^{\a\c} h^\b_\c + 2 \br_{\a\b\c\d}h^{\a\c} h^{\b\d}).
\ena
We use the notation $\simeq$ to denote equality up to total derivatives.

\section{Complete formulae for quadratic terms}
\label{complete}

Expanding the action~\p{action} around the background up to the second order,
we find
\bea
{\cal L} \!\! &=&\!\! \sqrt{-\bg} [ A+ 
 B + 
 C ],
\ena
where
\bea
A \hs{-2}&=&\hs{-2} \frac{\s}{\kappa^2} \br -2 \Lambda + \a \br^2+\b \br_{\mu\nu}^2
+ \c \br_{\mu\nu\rho\la}^2, \nn
B \hs{-2}&=&\hs{-2} \frac{\s}{\kappa^2} \Big(R^{(1)}+\frac{h}{2} \br\Big) -\Lambda h
+ \a \Big(\frac{h}{2} \br+2 R^{(1)} \Big) \br
+\b \Big[\frac{h}{2} \br_{\mu\nu}^2 +2 (\br^{\mu\nu} R^{(1)}_{\mu\nu}
-\br_{\mu\nu} \br^\mu_\a h^{\nu\a}) \Big] \nn
&&\hs{-3} +\c \Big[\br_{\mu\nu\rho\la}^2 \frac{h}{2}
+2 \br^{\mu\nu\rho\la} R_{\mu\nu\rho\la}^{(1)}
- 2 \br_{\mu\nu\rho\la}\br_\a{}^{\nu\rho\la} h^{\mu\a} \Big], \nn
C \hs{-2}&=&\hs{-2} \frac{\s}{\kappa^2} \Big[ R^{(2)} + R^{(1)}\frac{h}{2}
 +\frac{\br}{8}(h^2-2 h_{\mu\nu}^2) \Big] -\Lambda \frac{h^2-2 h_{\mu\nu}^2}{4} \nn
&&\hs{-3}
+\; \a \Big[ R^{(1) 2} + \br \Big( 2 R^{(2)}+ R^{(1)} h +\br \frac{h^2-2h_{\mu\nu}^2}{8}
\Big) \Big]
+ \b \Big[R_{\mu\nu}^{(1) 2} +2 \br^{\mu\nu} R^{(2)}_{\mu\nu}
-4 \br^{\mu\nu} R^{(1)}_{\mu\a} h_\nu^\a \nn
&&\hs{-3}
+\; 2 \br_{\mu\a}\br_\nu^\a h^{\mu\b}h_\b^\nu
+ \br_{\mu\nu}\br_{\a\b} h^{\mu\a}h^{\nu\b}
+(\br^{\mu\nu} R_{\mu\nu}^{(1)}-\br_{\mu\a} \br^\a_\nu h^{\mu\nu}) h
+\br_{\mu\nu}^2 \frac{h^2-2 h_{\mu\nu}^2}{8} \Big] \nn
&& \hs{-3} + \c \Big[ \br_{\mu\nu\rho\la} \br_\a{}^{\nu\rho\la}(h^2)^{\mu\a}
+ 2 \br_{\mu\a\rho\la}\br _{\nu\b}{}^{\rho\la} h^{\mu\nu} h^{\a\b}
- 4 \br^{\nu\rho\la}{}_\mu R_{\nu\rho\la\a}^{(1)} h^{\mu\a}
+2 \br^{\mu\nu\rho\la} R_{\mu\nu\rho\la}^{(2)} \nn
&& + R_{\mu\nu\rho\la}^{(1)\; 2} +( \br^{\mu\nu\rho\la} R_{\mu\nu\rho\la}^{(1)}
- \br_{\mu\nu\rho\la}\br_\a{}^{\nu\rho\la} h^{\mu\a})h
+ \frac{h^2-2 h_{\mu\nu}^2}{8} \br_{\mu\nu\rho\la}^2 \Big],
\ena
where $R_{\mu\nu}^{(1,2)}$ and $R^{(1,2)}$ are defined in appendix~\ref{conv}.

In our main text, we presented only quadratic terms in the fluctuation of
the curvature tensors relevant for our final results.
Here we give complete expressions for completeness.
Before doing any integration by parts, we find that the second variations are
\bea
\label{varalpha1}
\alpha
\hs{-5}
&& \hs{-2}
\Big[
\Box h\Box h
-2\Box h\nabla_\a\nabla_\b h^{\a\b}
+\nabla_\mu\nabla_\nu h^{\mu\nu}\nabla_\a\nabla_\b h^{\a\b}
\nn
&&
-\frac{1}{2}\br\nabla_\rho h\nabla^\rho h
-2\br \nabla_\mu h^{\mu\nu}\nabla_\a h^a{}_\nu
+2\br\nabla_\b h\nabla_\a h^{\a\b}
-\br \nabla_\a h^{\mu\nu}\nabla_\mu h^\a{}_\nu
+\frac{3}{2}\br \nabla^\rho h^{\mu\nu}\nabla_\rho h_{\mu\nu}
\nn
&&
+2\br h^{\mu\nu} \nabla_\mu\nabla_\nu h
-2\br h^{\mu\nu} \nabla_\mu\nabla_\a h^\a{}_\nu
-2\br h^{\mu\nu} \nabla_\a\nabla_\mu h^a{}_\nu
+\br h \nabla_\a\nabla_\b h^{\a\b}
\nn
&&
+2\br h^{\mu\nu}\Box h_{\mu\nu}
-\br h\Box h
-2 \br_{\mu\nu}h^{\mu\nu}\nabla_\a\nabla_\b h^{\a\b}
+2\br_{\mu\nu}h^{\mu\nu}\Box h
\nn
&&
+h^{\mu\nu}\br_{\mu\a}\br_{\nu\b}h^{\a\b}
+2 h^{\mu\nu}\br_{\mu\a}\br h^\a{}_\nu
-h\br\br_{\a\b}h^{\a\b}
+\left(\frac{1}{8}h^2-\frac{1}{4}h_{\mu\nu}h^{\mu\nu}\right)\br^2
\Big],
\ena
for scalar curvature squared,
\bea
\label{varbeta1}
\beta \hs{-5}
&& \hs{-2}
\Big[
\frac{1}{4}\nabla_\rho\nabla_\sigma h\nabla^\rho\nabla^\sigma h
-\nabla_\a\nabla_\rho h\nabla_\b\nabla^\rho h^{\a\b}
+\frac{1}{2}\nabla_\a\nabla_\b h\Box h^{\a\b}
+\frac{1}{4}\Box h^{\mu\nu}\Box h_{\mu\nu}
\nn
&&
-\nabla_\mu\nabla_\a h^{\mu\nu}\Box h^\a{}_\nu
+\frac{1}{2}\nabla_\mu\nabla_\a h^{\mu\nu}\nabla_\b\nabla_\nu h^{\a\b}
+\frac{1}{2}\nabla_\mu\nabla_\rho h^{\mu\nu}\nabla_\a\nabla^\rho h^\a{}_\nu
\nn
&&
+\frac{1}{2}\br^{\rho\sigma}\nabla_\rho h^{\mu\nu}\nabla_\sigma h_{\mu\nu}
+\br_\b{}^\rho\nabla_\a h\nabla_\rho h^{\a\b}
-\frac{1}{2}\br_{\a\b}\nabla_\rho h\nabla^\rho h^{\a\b}
-2\br_\mu{}^\rho\nabla_\rho h^{\mu\nu}\nabla_\a h^\a{}_\nu
\nn
&&
+\br^{\mu\nu}\nabla_\b h^{\mu\nu}\nabla_\a h^{\a\b}
-\br_{\mu\a} \nabla_\b h^{\mu\nu}\nabla_\nu h^{\a\b}
+\br_{\mu\a}\nabla_\rho h^{\mu\nu}\nabla^\rho h^\a{}_\nu
\nn
&&
+h^{\mu\nu}\br_{\a\b}\nabla_\mu\nabla_\nu h^{\a\b}
+2h^{\mu\nu}\br_\mu{}^\rho\nabla_\rho\nabla_\nu h
-2h^{\mu\nu}\br_{\mu\b}\nabla_\a\nabla_\nu h^{\a\b}
-2h^{\mu\nu}\br_\mu{}^\rho\nabla_\a\nabla_\rho h^\a{}_\nu
\nn
&&
+h^{\mu\nu}\br^{\rho\sigma}\nabla_\rho\nabla_\sigma h_{\mu\nu}
+2 h^{\mu\nu}\br_{\mu\a}\Box h^\a{}_\nu
-2h^{\mu\nu}\br_\a{}^\rho\nabla_\mu\nabla_\rho h^\a{}_\nu
\nn
&&
+h\br_\a{}^\rho\nabla_\b\nabla_\rho h^{\a\b}
-\frac{1}{2}h\br_{\a\b}\Box h^{\a\b}
-\frac{1}{2}h\br^{\rho\sigma}\nabla_\rho\nabla_\sigma h
+h^{\mu\nu}\br_{\mu\a}\br_{\nu\b}h^{\a\b}
\nn
&&
+2h^{\mu\nu}\br_{\mu\rho}\br^\rho{}_\a h^\a{}_\nu
-h \br_{\a\rho}\br^\rho{}_\b h^{\a\b}
+\left(\frac{1}{8}h^2-\frac{1}{4}h_{\mu\nu}h^{\mu\nu}\right)
\br_{\rho\s}\br^{\rho\s}
\Big],
\ena
for Ricci curvature squared, and
\bea
\label{vargamma1}
\c \hs{-5}
&& \hs{-2}
\Big[
\nabla_\a\nabla_\b h^{\mu\nu}\nabla_\mu\nabla_\nu h^{\a\b}
-\nabla_\rho\nabla_\a h^{\mu\nu}\nabla^\rho\nabla_\mu h^\a{}_\nu
-\nabla_\a\nabla_\rho h^{\mu\nu}\nabla_\mu \nabla^\rho h^\a{}_\nu
+\nabla_\rho\nabla_\sigma h^{\mu\nu}\nabla^\rho\nabla^\sigma h_{\mu\nu}
\nn
&&
+4\br_{\mu\a\nu\rho}\nabla_\beta h^{\mu\nu}\nabla^\rho h^{\a\b}
-\br_{\mu\a\nu\b}\nabla_\rho h^{\mu\nu}\nabla^\rho h^{\a\b}
-2\br_{\mu\a\rho\sigma}\nabla^\rho h^{\mu\nu}\nabla^\sigma h^\a{}_\nu
+2\br_{\mu\rho\a\sigma}\nabla^\sigma h^{\mu\nu}\nabla^\rho h^\a{}_\nu
\nn
&&
+h^{\mu\nu}\br_{\mu\rho\a\sigma}(4\nabla^\rho\nabla^\sigma+2\nabla^\sigma\nabla^\rho)h^\a{}_\nu
+h^{\mu\nu}\br_{\mu\a\rho\b}(4\nabla_\nu\nabla^\rho+2\nabla^\rho\nabla_\nu)h^{\a\b}
-2h\br_{\a\rho\b\sigma}\nabla^\sigma\nabla^\rho h^{\a\b}
\nn
&&
+\frac52h^{\mu\nu}h^\a{}_\nu \br_{\mu\lambda\rho\sigma}\br_\a{}^{\la\rho\sigma}
+\frac{1}{2}h^{\mu\nu}h^{\a\b}\br_{\mu\a\rho\sigma}\br_{\nu\b}{}^{\rho\sigma}
-h h^{\a\b}\br_{\a\lambda\rho\sigma}\br_\b{}^{\la\rho\sigma}
\nn
&&
+\left(\frac{1}{8}h^2-\frac{1}{4}h_{\mu\nu}h^{\mu\nu}\right)
\br_{\rho\s\la\tau}\br^{\rho\s\la\tau}
\Big],
\ena
for Riemann curvature squared.

We now integrate by parts derivatives in order to write the variations
as $h{\cal O}h$ where ${\cal O}$ is a fourth order operator.
There is some arbitrariness in the presentation of the formula,
due to the freedom of performing commutations and integrations by parts.
In order to reduce it, let us make some conventions.
We put $h^{\mu\nu}$ on the left and $h^{\alpha\beta}$ on the right.
It is convenient to put $\nabla_\alpha$ and $\nabla_\beta$ on the right
and $\nabla_\mu$ and $\nabla_\nu$ on the left,
so that they form the vector combination $h^\mu$ whenever possible.
Also, we use the convention that in those terms where only one of the $h$'s is traced,
it stays on the left.

After some manipulations, (\ref{varalpha1}) can be rewritten in the form
\bea
&&
\a h^{\mu\nu} \Big[ \nabla_\mu \nabla_\nu \nabla_\a \nabla_\b
- 2\bg_{\mu\nu} \Box \nabla_\a \nabla_\b 
+\bg_{\mu\nu} \bg_{\a\b} \Box^2
\nn
&&
-\bg_{\nu\b}\br\nabla_\mu \nabla_\a
-2\br_{\mu\nu}\nabla_\a \nabla_\b
+\bg_{\mu\nu}\br\nabla_\a \nabla_\b
+2\bg_{\mu\nu}\br_{\a\b}\Box
+\frac12(\bg_{\mu\a}\bg_{\nu\b}-\bg_{\mu\nu}\bg_{\a\b})\br \Box
\nn
&&
-\bg_{\mu\nu}\br\br_{\a\b}
-\frac14J_{\mu\nu\a\b}\br^2
+\bg_{\nu\b}\br\br_{\mu\a}
+ \br_{\mu\nu} \br_{\a\b}
+\br \br_{\mu\a\nu\b}
\nn
&&
+2\bg_{\mu\nu}\Box \br_{\a\b}
+2\bg_{\mu\nu}\nabla_\a\nabla_\b \br
-\bg_{\nu\beta}\nabla_\mu\nabla_\a \br
+\frac14 \left(3\bg_{\mu\a}\bg_{\nu\b}+\bg_{\mu\nu}\bg_{\a\b}\right)\Box\br
\nn
&&
+\bg_{\nu\b}\nabla_\mu\br\nabla_\a
+4\bg_{\mu\nu}\nabla_\rho \br_{\alpha\beta} \nabla^\rho
+2\bg_{\mu\nu}\nabla_\a \br\nabla_\b
\Big] h^{\a\b},
\label{varalpha2}
\ena
where $J$ is given by (\ref{defJ}).
We have checked that this formula agrees with the $\alpha$--terms
in Eq.~(3.11) or (3.15) of \cite{BC}.
Dropping the last two lines, this agrees with (\ref{r1}) up to
some integrations by parts and commutations of derivatives.

Similarly one can rewrite (\ref{varbeta1}) in the ``standard'' form:
\bea
\label{varbeta2}
\beta h^{\mu\nu}\hs{-7}&&
\Big[\frac{1}{2}\nabla_\mu \nabla_\nu \nabla_\a \nabla_\b
-\frac{1}{2}\bg_{\mu\nu} \Box \nabla_\a \nabla_\b
-\frac{1}{2}\bg_{\nu\b} \nabla_\mu \Box \nabla_\a
+\frac14( \bg_{\mu\a} \bg_{\nu\b}+\bg_{\mu\nu} \bg_{\a\b})\Box^2
\\
&&
+\frac{1}{2}\br_{\mu\a} \nabla_\nu \nabla_\b
-2 \bg_{\nu\b} \br_\mu^\rho  \nabla_\rho\nabla_\a
+ \bg_{\mu\nu}\br_\a^\rho \nabla_\rho \nabla_\b
+ \br_{\mu\a\nu\b} \Box
+\frac{1}{2}J_{\mu\nu\a\b}\br^{\rho\la}\nabla_\rho \nabla_\la
\nn
&&
+\frac{1}{2}\bg_{\nu\b} \br_{\mu\rho}\br^{\rho}_\a
+\frac{1}{2}\br_{\mu\a} \br_{\nu\b}
+ \br^\rho_\mu \br_{\rho\a\nu\b}
- \bg_{\mu\nu}\br^{\rho\sigma} \br_{\rho\a\sigma\b}
+ \br_{\rho\mu\sigma\nu} \br^\rho{}_\a{}^\sigma{}_\b
-\frac14 J_{\mu\nu\a\b}\br_{\rho\la}\br^{\rho\la}
\nn
&&
+\frac{1}{2}\bg_{\mu\nu}\Box \br_{\a\b}
+\frac{1}{2}\bg_{\mu\nu}\nabla_\alpha\nabla_\beta \br
+\frac18 J_{\mu\nu\a\b}\Box \br
+2\bg_{\nu\beta}\nabla_\mu \br_{\alpha\rho}\nabla^\rho
-\bg_{\nu\beta}\nabla_\rho \br_{\mu\alpha}\nabla^\rho
\nn
&&
+\nabla_\alpha \br_{\mu\nu}\nabla_\beta
-\frac{1}{2}\nabla_\mu \br_{\nu\beta}\nabla_\alpha
+(\nabla_\alpha \br_{\mu\beta}-\nabla_\mu \br_{\alpha\beta})\nabla_\nu
+\frac{1}{2}\bg_{\mu\nu}\nabla_\alpha \br\nabla_\beta
+\bg_{\mu\nu}\nabla_\rho \br_{\alpha\beta}\nabla^\rho
\Big] h^{\a\b}.
\nonumber
\ena
This formula agrees exactly with the $\beta$--terms in Eq.~(3.11) or (3.15) of \cite{BC}.
Notice that if one neglects the terms of the form $\nabla\nabla\br$ and $\nabla\br\nabla$
(the last two lines), then there is some ambiguity in the form of the $\br^2$ terms,
because there is some combination of $\br^2$ terms that can be rewritten in
the form of the terms of the last two lines.
More precisely
\begin{equation}
\label{combo1}
[\nabla_\rho,\nabla_\mu]\br^\rho{}_\nu=\br_{\mu\rho}\br^\rho{}_\nu
-\br^{\rho\sigma}\br_{\mu\rho\nu\sigma}.
\end{equation}
If one subtracts this expression from (\ref{varbeta2}), the fourth term in the
third row is replaced by $-g_{\mu\nu}\br_\alpha{}^\rho\br_{\rho\beta}$.
Dropping the last two lines and performing some commutations of derivatives
and integrations by parts, one obtains equation (\ref{r2}).

Finally we consider the terms proportional to $\c$.
Now we encounter products of two Riemann tensors.
Due to $\br_{[\mu\nu\rho]\sigma}=0$, there are various ways of writing these products.
We have
\begin{equation}
\br_{\mu\rho\a\sigma}\br_\nu{}^\sigma{}_\beta{}^\rho=
\br_{\mu\rho\a\sigma}\br_\nu{}^\rho{}_\beta{}^\sigma
-\br_{\mu\rho\a\sigma}\br_{\nu\beta}{}^{\rho\sigma}
\end{equation}
Furthermore, when contracted with $h^{\mu\nu}h^{\a\b}$ we can replace
\begin{equation}
\br_{\mu\rho\nu\sigma}\br_\a{}^\sigma{}_\beta{}^\rho\leftrightarrow
\br_{\mu\rho\nu\sigma}\br_\a{}^\rho{}_\beta{}^\sigma\ ;
\qquad
2\br_{\mu\rho\a\sigma}\br_{\nu\beta}{}^{\rho\sigma}\leftrightarrow
\br_{\mu\a\rho\sigma}\br_{\nu\beta}{}^{\rho\sigma}\ .
\end{equation}
Using these properties we choose the following basis of independent combinations:
\bea
\br_{\mu\rho\a\sigma} \br_\nu{}^\rho{}_\beta{}^\sigma\ ;\ \
\br_{\mu\a\rho\sigma}\br_{\nu\b}{}^{\rho\sigma} \ ;\ \
\br_{\mu\rho\nu\sigma} \br_\a{}^\rho{}_\b{}^\sigma\ ;\ \
\br_{\mu\rho}\br_{\nu\a}{}^\rho{}_\b
\ena
After integrations by parts and arranging in canonical order, (\ref{vargamma1}) becomes
\bea
\c h^{\mu\nu} \hs{-5}&& \hs{-2} \Big[
\nabla_\mu \nabla_\nu \nabla_\a \nabla_\b
-2 \bg_{\nu\b}\nabla_\mu \Box \nabla_\a
+ \bg_{\mu\a} \bg_{\nu\b} \Box^2
+ 3 \br_{\mu\a\nu\b}\Box
-2\bg_{\nu\b}\br_{\mu\a}\Box
-2\bg_{\mu\nu} \br_{\a\rho\b\sigma}\nabla^\rho\nabla^\sigma
\nn
&&
+2 \bg_{\nu\b} \br_{\mu\rho\a\sigma}\nabla^\rho\nabla^\sigma
+4\br_{\a\mu\rho\nu} \nabla^\rho \nabla_\b
+ 4 \br_{\mu\a} \nabla_\nu \nabla_\b
-4 \bg_{\nu\b} \br_{\mu\rho}\nabla^\rho\nabla_\a
+ \bg_{\mu\a} \bg_{\nu\b} \br_{\rho\sigma} \nabla^\rho \nabla^\sigma
\nn
&&
-\bg_{\mu\nu}\br_{\a\lambda\rho\sigma}\br_\b{}^{\lambda\rho\sigma}
+2\br_{\mu\a} \br_{\nu\b}
-2\bg_{\nu\b}\br_{\mu\rho} \br^\rho{}_\a
+2\bg_{\nu\b}\br_{\mu\la\rho\sigma}\br_\a{}^{\la\rho\sigma}
-\frac{1}{4} J_{\mu\nu\a\b}\br_{\rho\s\la\tau}\br^{\rho\s\la\tau}
\nn
&&
+5\br_{\mu\rho\a\sigma} \br_\nu{}^\rho{}_\beta{}^\sigma
-4\br_{\mu\a\rho\sigma}\br_{\nu\b}{}^{\rho\sigma}
-3\br_{\mu\rho\nu\sigma} \br_\a{}^\rho{}_\b{}^\sigma
+3\br_{\mu\rho}\br_{\nu\a}{}^\rho{}_\b
\nn
&&
+8(\nabla_\a\br_{\mu\nu}-\nabla_\mu\br_{\a\nu})\nabla_\b
+8\nabla_\a(\nabla_\b\br_{\mu\nu}-\nabla_\mu\br_{\b\nu})
+3\nabla_\rho\br_{\mu\a\nu\b}\nabla^\rho
+2\nabla_\a\br_{\b\mu\rho\nu}\nabla^\rho
\nn
&&
+2\bg_{\nu\b}\nabla_\mu\br_{\a\rho}\nabla^\rho
-4\bg_{\nu\b}\nabla_\rho\br_{\a\mu}\nabla^\rho
+\frac{1}{2}\bg_{\mu\a}\bg_{\nu\b}\nabla_\rho\br\nabla^\rho
\Big] h^{\a\b}.
\label{vargamma2}
\ena
If one neglects terms of the type in the last two lines,
then the coefficients of the terms in the fourth line can differ.
One can add an arbitrary multiple of the combination
\bea
[\nabla_\rho,\nabla_\mu]\br_{\nu\a}{}^\rho{}_\b=
\br_{\mu\rho\a\sigma} \br_\nu{}^\rho{}_\beta{}^\sigma
-\br_{\mu\a\rho\sigma}\br_{\nu\b}{}^{\rho\sigma}
-\br_{\mu\rho\nu\sigma} \br_\a{}^\rho{}_\b{}^\sigma
+\br_{\mu\rho}\br_{\nu\a}{}^\rho{}_\b
\ena
If one subtracts six times this combination from (\ref{vargamma2})
and adds $g_{\nu\beta}$ times a combination of the type (\ref{combo1}),
and uses the identity
\bea
\br_\mu{}^{\rho\la\s} \br_{\a\s\rho\la}
= -\frac12 \br_\mu{}^{\rho\la\s}\br_{\a\rho\la\s},
\label{cid}
\ena
then one sees that (\ref{r3}), written in the basis given above,
agrees with (\ref{vargamma2}) with the last two lines removed.

\section{$U$ and $V$}
\label{uv}

The tensors $U$ and $V$ defined in (\ref{hami}) are obtained
by acting with $K^{-1}$ on the tensors $W$ and $D$ given in (\ref{curm}) and (\ref{der}).
We have
\bea
(K^{-1})_{\mu\nu}{}^{\a\b}
=\frac{4}{\b+4\c} (\d_{\mu\nu}{}^{\a\b} -\Omega \bg_{\mu\nu} \bg^{\a\b}),
\ena
with
\bea
\Omega = \frac{4\a+\b}\Sigma, ~~~
\Sigma \equiv 4(\c-\a)+D(4\a+\b).
\label{sigma}
\ena
After some work we find
\bea
(U)_{\mu\nu,\a\b}\!\!&=&\!\! \frac{4}{\b+4\c} \Bigg[
\frac32 \c \bg_{\nu\b}\br_\mu{}^{\rho\la\s} \br_{\a\rho\la\s}
-\c \br^\la{}_{\a\mu}{}^\rho \br_{\la\nu\b\rho}
+4\c \br_{\rho\a\mu\la} \br_{\nu\b}{}^{\rho\la} \nn
&& -3\c(\br_\mu^\s \br_{\s\a\nu\b} + \br^\s_\a \br_{\s\mu\b\nu})
+\Big(\frac{\b}{2}+\c \Big) \br_{\mu\a}\br_{\nu\b}
-\frac{\c}{2}\bg_{\a\b} \br_{\mu\rho\la\s} \br_\nu{}^{\rho\la\s} \nn
&& +\frac14 S^2 (\Omega_1 \bg_{\mu\nu} \bg_{\a\b} -\bg_{\mu\a} \bg_{\nu\b} )
+\Big(\frac{\a}{2}\br +\frac{\s}{4\kappa^2} \Big)
 (\br_{\mu\a\nu\b}+ 3\bg_{\nu\b} \br_{\mu\a} -\bg_{\a\b} \br_{\mu\nu})\nn
&& +\Big(\frac{5}{2}\b+4\c \Big) \bg_{\nu\b} \br_{\mu\s} \br^{\s}_\a
+(\b+5\c) \br_{\rho\mu\la\nu} \br^\rho{}_\a{}^\la{}_\b
- \frac{\b}{2} \bg_{\a\b} \br_{\mu\s}\br^\s_\nu \nn
&& - \c \Omega_1 \bg_{\mu\nu} \br_{\a\rho\la\s} \br_\b{}^{\rho\la\s}
- \b \Omega_1 
 \bg_{\mu\nu} \br_{\a\s}\br^\s_\b
+ \a \br_{\mu\nu} \br_{\a\b}
-\Big( \frac{\s}{\kappa^2} \Omega_3 +\a \Omega_1 \br \Big) \bg_{\mu\nu} \br_{\a\b} \nn
&& -\frac{1}{4\kappa^2}(\s\br-4\Lambda)\Omega \bg_{\mu\nu} \bg_{\a\b}
-(\b+4\c) \bg_{\nu\b} \br^{\rho\la}\br_{\mu\rho\a\la} \Bigg],
\ena
where we have defined
\bea
S^2 = \a \br^2+\b \br_{\mu\nu}^2 + \c \br_{\mu\nu\rho\la}^2
+ \frac{1}{\kappa^2}(\s \br-2 \Lambda),
\ena
and
\bea
\Omega_1 = \frac{10\a+3\b+2\c}{\Sigma}, ~~~
\Omega_3 = \frac{3\a+\b+\c}{\Sigma}, ~~~
\ena
with $\Sigma$ given in \p{sigma}.

We also find the expression for $V^{\rho\la}=(V^{\rho\la})_{\mu\nu,\a\b}$ is
\newcommand{\bk}{{\bf k}}
\bea
V^{\rho\la} = \frac{4}{\b+4\c} \sum_{i=1}^{20} b_i \bk_i,
\ena
where
\bea
&& \bk_1= \bg_{\nu\b} \bg^{\rho\la} \br_{\mu\a},~~~
\bk_2= \d_{\mu\nu,\a\b} \bg^{\rho\la},~~~
\bk_3= \bg^{\rho\la} \br_{\mu\a\nu\b},~~~
\bk_4= \d_{\nu\b}{}^{\rho\la} \br_{\mu\a}, \nn
&& \bk_5= \d_{\nu\b}{}^{\rho\la} \bg_{\mu\a},~~~~
\bk_6= \d_{\mu\nu,\a\b}\br^{\rho\la},~~~
\bk_7= \frac12 (\d_\nu^{(\rho} \br^{\la)}{}_{\a\b\mu} + \d_\b^{(\rho} \br^{\la)}{}_{\mu\nu\a}),\nn
&& \bk_8= \bg_{\nu\b}\d^{(\rho}_{(\mu} \br^{\la)}_{\a)}, ~~~
\bk_9 = \bg_{\nu\b} \br_{(\a}{}^{\rho\la}{}_{\mu)},~~~
\bk_{10}= \frac12 (\d_{\a\b}{}^{\rho\la}\br_{\mu\nu} + \d_{\mu\nu}{}^{\rho\la} \br_{\a\b}),\nn
&& \bk_{11} = \bg_{\mu\nu} \br_{\a}{}^{\rho\la}{}_{\b},~~~
\bk_{12} = \bg_{\a\b} \br_{\mu}{}^{\rho\la}{}_{\nu},~~~
\bk_{13} = \bg_{\mu\nu} \bg^{\rho\la} \br_{\a\b},~~~
\bk_{14} = \bg_{\a\b} \bg^{\rho\la} \br_{\mu\nu}, \nn
&& \bk_{15} = \bg_{\mu\nu} \d_\a^\la \br_\b^\rho,~~~~
\bk_{16} = \bg_{\a\b} \d_\mu^\la \br_\nu^\rho,~~~
\bk_{17}= \bg_{\mu\nu} \d_{\a\b}{}^{\rho\la},~~~
\bk_{18}= \bg_{\a\b} \d_{\mu\nu}{}^{\rho\la}, \nn
&& \bk_{19} = \bg_{\mu\nu} \bg_{\a\b} \bg^{\rho\la}, ~~~
\bk_{20} = \bg_{\mu\nu} \bg_{\a\b} \br^{\rho\la},
\ena
and
\bea
&& b_1= -2\c, ~~~
b_2=\frac{\a}{2} \br +\frac{\s}{4\kappa^2}, ~~~
b_3 = \b+3\c, ~~~
b_4 = 2\c, ~~~
b_5 = -\frac{\s}{2\kappa^2} -\a \br,\nn
&& b_6 = \frac{\b}{2}+\c,~~~
b_7 = -4\c, ~~~
b_8 = -2\b-4\c, ~~~
b_9 = -2\c, ~~~
b_{10} = -2\a, \nn
&& b_{11} = 4\c \Omega_3,~~~
b_{12} = \c, ~~~
b_{13} = -\b \Omega_3, ~~~
b_{14} = \a, ~~~
b_{15} = 2\b \Omega_3,~~~
b_{16} = \frac{\b}{2}, \nn
&& b_{17} = 2\a \Omega_3 \br +\s \frac{\Omega_1-2\Omega}{2\kappa^2}, ~~~
b_{18} = \frac{\a}{2}\br+\frac{\s}{4\kappa^2}, ~~~
b_{19}=-b_{17}, ~~~
b_{20}=-\b \Omega_3.
\label{v}
\ena
This result agrees with \cite{BS2} after the replacement $\s\to-1$ and $\Lambda\to -\Lambda$,
which corresponds to Euclideanization made in \cite{BS2}.

\section{Traces}
\label{traces}

The trace of $U$ is given as
\bea
\mbox{tr }U = \d^{\mu\nu,\a\b} U_{\mu\nu,\a\b}
= A_1 \br_{\mu\nu\rho\la}^2 + A_2 \br_{\mu\nu}^2
+ A_3 \br^2+A_4 \frac{\s\br}{\kappa^2}+A_5 \frac{\Lambda}{\kappa^2},
\ena
where
\bea
A_k = \frac{p_{k0}+p_{k1} D +p_{k2} D^2+ p_{k3} D^3}{2(\b+4\c)\Sigma},~~~
(k=1,2,3,4,5),
\ena
and
\bea
&& p_{10}=-8(3\a\b+20\a\c-8\c^2),~~~
p_{11}=2 (12 \a\b + 3\b^2 +40\a\c+13\b\c+12\c^2), \nn
&& p_{12}=(24\a+5\b-4\c)\c,~~~
p_{13}=-\c(4\a+\b),~~~
p_{20}=-8(4\a^2+12\a\b+3\b^2+8\a\c-12\c^2), \nn
&&
p_{21}= 2[ 16 \a^2+12\a(\b+4\c)+\b(5\b+24\c)],~~~
p_{22}=(24\a+5\b-4\c)\b,~~~
p_{23}=-\b(4\a+\b),\nn
&&
p_{30}=-8[10\a^2+4\a(\b+\c)-(\b+2\c)\c],~~~
p_{31}=2(7\a\b+\b^2+20\a\c+ 2 \b\c),\nn
&& p_{32}=(24\a+5\b-4\c)\a,~~~
p_{33}=-\a(4\a+\b),~~~
p_{40}=-8(3\a+\b+\c), \nn
&& p_{41}=4(\a+\b+3\c),~~~
p_{42}=2(6\a+\b-2\c),~~~
p_{43}=-(4\a+\b),~~~
p_{50}=0, \nn
&& p_{51}=-4(4\a+\b),~~~
p_{52}=2(\b+4\c),~~~
p_{53}=2(4\a+\b).
\ena

Next,
\bea
\mbox{tr } (V^\rho_\rho \br)
= B_1 \br^2 + B_2 \frac{\s\br}{\kappa^2},
\ena
where
\bea
B_k = \frac{l_{k0}+l_{k1} D +l_{k2} D^2+ l_{k3} D^3+l_{k4}D^4}{2(\b+4\c)\Sigma},~~~
(k=1,2),
\ena
and
\bea
&& l_{10} = 16(2\a+\b-4\c)(2\a+\b+2\c),~~
l_{11} = -4 (12 \a^2 + 6 \a \b + 5 \b^2 - 4 \a \c + 20 \b \c + 16 \c^2),\nn
&& l_{12} = - 2[4 \a^2 + 28 \a \b + 5 \b^2 + 4 (9 \a + \b) \c + 8 \c^2],~~
l_{13} = 2 [-8 \a^2 + 3 \a \b + \b^2 - 2 (2 \a + \b) \c], \nn
&& l_{14} = 2 \a (4 \a + \b),~~
l_{20}=0, ~~
l_{21} = 4 (2 \a + \b + 2 \c),~~
l_{22} = - 4 (\a + \b + 3 \c) ,\nn
&& l_{23} = - (8 \a + \b - 4 \c) , ~~
l_{24} = (4 \a + \b) .
\ena
\bea
\mbox{tr } (V^{\rho\la} \br_{\rho\la})
= C_1 \br_{\mu\nu}^2 + C_2 \br^2+C_3 \frac{\s\br}{\kappa^2},
\ena
where
\bea
C_k &=& \frac{q_{k0}+q_{k1} D +q_{k2} D^2+ q_{k3} D^3}{(\b+4\c)\Sigma},~~~
(k=1,2), \nn
C_3 &=& \frac{(D-1)(D-2)[4(D+1)\a+(D+2)\b+4\c]}{2(\b+4\c)\Sigma},
\ena
and
\bea
&&\hs{-5} q_{10} = 8(2\a+\b-4\c)(2\a+\b+2\c),~~
q_{11} = -2 (16 \a^2 + 8 \a \b + 3 \b^2 +20 \a \c + 14 \b \c + 4 \c^2),\nn
&&\hs{-5}
q_{12} = - 16 \a(\b+\c) - (\b-2\c)(3\b+4\c),~~
q_{13} = (4\a+\b)(\b+2\c), \nn
&& \hs{-5}
q_{20}=4 [2 \a^2 - \b^2 - 3 \b \c - 6 \c^2 + \a (\b + 12 \c)],~~
q_{21} = -2 (2 \a^2 + 6 \a \b + \b^2 + 10 \a \c + 3 \b \c + 8 \c^2),\nn
&&\hs{-5} q_{22} = - [8 \a^2 + 4 \b \c + \a (\b + 12 \c)],~~
q_{23} = \a (4 \a + \b).
\ena
The following results are obtained using software Math Tensor run on the Mathematica.
It is important to realize that the indices are symmetrized. For example,
the indices $\rho$ and $\la$ on $V$ must be symmetrized in making products.
\bea
\frac{1}{48} \mbox{tr } (V^\rho_\rho V^\la_\la)
+\frac{1}{24} \mbox{tr } (V_{\rho_\la} V^{\rho\la})
=D_1 \br_{\mu\nu\rho\la}^2 + D_2 \br_{\mu\nu}^2+D_3 \br^2+D_4 \frac{\s\br}{\kappa^2}
+D_5 \frac{\s^2}{\kappa^4},
\ena
where
\bea
D_1 &=& \frac{q_{10}+q_{11} D+ q_{12} D^2+ q_{13} D^3}{96(\b+4\c)^2 \Sigma}, \nn
D_i &=& \frac{q_{i0}+q_{i1} D+ q_{i2} D^2+ q_{i3} D^3+q_{i4}D^4+q_{i5}D^5+q_{i6}D^6}
{96(\b+4\c)^2 \Sigma^2}, ~~ (i=2,\cdots, 5),
\ena
with
\bea
q_{10} &=& -1536 \c [-2 \c^2 + \a (\b + 6 \c)],~~
q_{11} = -192 [\a (\b^2 + 2 \b \c - 6 \c^2) - \c (3 \b^2 + 18 \b \c + 26 \c^2)], \nn
q_{12} &=& 48 [\b^3 + 12 \b^2 \c + 36 \b \c^2 + 18 \c^3 + 2 \a (\b^2 + 14 \b \c + 39 \c^2)],\nn
q_{13} &=& 24 (4 \a + \b) (\b + 3 \c)^2, \\
q_{20} &=& 256 [8 \a^4 + \b^4 - 32 \a^3 \c + 4 \b^3 \c - 2 \b^2 \c^2 - 8 \c^4 +
    10 \a^2 (\b + 4 \c)^2\nn
&&  - 4 \a \c (5 \b^2 + 28 \b \c + 16 \c^2)],\nn
q_{21} &=& -128 [32 \a^4 - \b^4 + 4 \a^3 (3 \b - 20 \c) - 6 \b^3 \c- 27 \b^2 \c^2
 - 84 \b \c^3 - 120 \c^4\nn
&& + \a^2 (33 \b^2 + 148 \b \c + 136 \c^2) + 2 \a (3 \b^3 + 27 \b^2 \c + 74 \b \c^2
 + 112 \c^3)], \nn
q_{22} &=& 16 [128 \a^4 + 15 \b^4 + 64 \b^3 \c + 360 \b^2 \c^2 + 704 \b \c^3 +
     448 \c^4 + 16 \a^3 (9 \b + 4 \c) \nn
&& + 4 \a^2 (29 \b^2 - 204 \b \c - 336 \c^2) + 4 \a (21 \b^3 + 2 \b^2 \c
 + 144 \b \c^2 + 256 \c^3)], \nn
q_{23} &=& 8 [5 \b^4 + 104 \b^3 \c + 344 \b^2 \c^2 + 448 \b \c^3 + 64 \c^4
 - 128 \a^3 (\b + 4 \c) \nn
&& + 8 \a^2 (\b^2 + 64 \b \c + 40 \c^2) + 4 \a (7 \b^3 +140 \b^2 \c+352 \b \c^2+416 \c^3)],\nn
q_{24} &=& 8 [32 \a^3 (\b + 4 \c) + 8 \a^2 (7 \b^2 + 60 \b \c + 96 \c^2)
 + 8 \a (3 \b^3 + 28 \b^2 \c + 52 \b \c^2 + 16 \c^3) \nn
&& + \b (3 \b^3 + 28 \b^2 \c + 56 \b \c^2 + 32 \c^3)],~~
q_{25} = 32 (4 \a + \b)^2 \c^2, ~~~~
q_{26} = 0, \\
q_{30} &=& 128 [8 \a^4 - \b^4 - 6 \b^3 \c - 18 \b^2 \c^2 - 32 \b \c^3 - 56 \c^4
 + 16 \a^3 (\b + 5 \c) \nn
&& + 6 \a^2 (\b^2 - 40 \c^2) + 2 \a (\b^3 + 12 \b^2 \c + 24 \b \c^2 + 72 \c^3)], \nn
q_{31} &=& -64 [36 \a^4 + 3 \b^4 + 13 \b^3 \c + 36 \b^2 \c^2 + 96 \b \c^3 +
     68 \c^4 + 8 \a^3 (8 \b + 21 \c) \nn
&& + 2 \a^2 (13 \b^2 - 8 \b \c - 108 \c^2) +
     \a (13 \b^3 + 10 \b^2 \c - 48 \b \c^2 + 72 \c^3)], \nn
q_{32} &=& 8 [160 \a^4 - \b^4 + 48 \a^3 (7 \b - 4 \c) - 44 \b^3 \c -
     204 \b^2 \c^2 - 304 \b \c^3 - 32 \c^4 \nn
& & + 16 \a^2 (9 \b^2 + 5 \b \c + 68 \c^2) -
     4 \a (\b^3 + 70 \b^2 \c + 196 \b \c^2 + 272 \c^3)],\nn
q_{33} &=& 4 [80 \a^4 + \a^3 (-64 \b + 544 \c) - 8 \a^2 (27 \b^2 + 56 \b \c + 174 \c^2) \nn
&& -   \b (7 \b^3 + 32 \b^2 \c + 100 \b \c^2 + 32 \c^3) - 4 \a (17 \b^3 + 52 \b^2 \c
 + 160 \b \c^2 + 48 \c^3)],\nn
q_{34} &=& \! -4 [80 \a^4 + 136 \a^3 \b + 18 \a^2 \b^2 - 6 \a \b^3 - \b^4 +
     4 \b (2 \a^2 + 6 \a \b + \b^2) \c \nn
&& + 4 (36 \a^2 + 16 \a \b + \b^2) \c^2], \nn
q_{35} &=& -4 \a (4 \a + \b) (4 \a^2 - 9 \a \b - 2 \b^2 + 8 \a \c + 4 \b \c),~~
q_{36} = 4 \a^2 (4 \a + \b)^2, \\
q_{40} &=& 128 [\b^3 + 6 \b^2 \c + 4 \b \c^2 + 8 \c^3 + 4 \a^2 (\b + 10 \c) -
     8 \a \c (\b + 10 \c)], \nn
q_{41} &=& -32 [8 \a^3 + \b^3 + 8 \b^2 \c - 16 \c^3 + 32 \a^2 (\b + 5 \c) +
       2 \a (\b^2 - 60 \c^2)],\nn
q_{42} &=& 16 [16 \a^3 + 48 \a^2 \b + 14 \a \b^2 - \b^3 -
     8 (10 \a^2 + 3 \a \b + 2 \b^2) \c + 8 (22 \a + \b) \c^2 - 16 \c^3], \nn
q_{43} &=& 8 [40 \a^3 - \b^3 + 160 \a^2 \c + 12 \b^2 \c - 16 \c^3 -
     8 \a (2 \b^2 - 2 \b \c + 15 \c^2)], \nn
q_{44} &=& -4 [80 \a^3 + 88 \a^2 \b + 18 \a \b^2 + \b^3 - 4 \b (10 \a + \b) \c +
       16 (\a + \b) \c^2], \nn
q_{45} &=& -4 (4 \a + \b) [4 \a^2 - 5 \a \b - \b (\b - 2 \c)], ~~~
q_{46} = 4 \a (4 \a + \b)^2, \\
q_{50} &=& 0, ~~
q_{51} = -16 (4 \a^2 - \b^2 - 8 \a \c - 8 \b \c - 12 \c^2) , \nn
q_{52} &=& 8 (8 \a^2 + 4 \a \b - \b^2 - 12 \b \c - 24 \c^2) ,~~
q_{53} = 8 (10 \a^2 - \b^2 - 20 \a \c - 8 \b \c - 6 \c^2) , \nn
q_{54} &=& -2 (40 \a^2 + 20 \a \b + \b^2 - 12 \b \c - 24 \c^2) , \nn
q_{55} &=& -(4 \a + \b) (4 \a - \b - 8 \c) , ~~
q_{56} = (4 \a + \b)^2 .
\ena

Though these results agree with Ref.~\cite{BS2} if we set $\s=-1$ and $\Lambda\to -\Lambda$,
which correspond to Euclideanization made there, we have included these
for completeness.

\end{document}